# Pros and cons of the technique of processing IRMS data as desired-δ values: uncertainty and comparability in results, a case study for determining carbon and oxygen isotopic abundance ratios as $CO_2^+$

*B. P. Datta* (email: bibek@vecc.gov.in)
Radiochemistry Laboratory, Variable Energy Cyclotron Centre, Kolkata 700 064, India

## ABSTRACT

In isotope ratio mass spectrometry (IRMS), any sample (***S***) measurement is performed as a relative-difference ($^{S/W}\boldsymbol{\delta}_i$) from a working-lab-reference (***W***), but the result is evaluated relative to a recommended-standard (***D***): $^{S/D}\boldsymbol{\delta}_i$. It is thus assumed that different source specific results ($^{S1/D}\boldsymbol{\delta}_i$, $^{S2/D}\boldsymbol{\delta}_i$ **…**) would represent their sources (***S*1**, ***S*2** **…**), and be accurately intercomparable. However, the assumption has never been checked. In this manuscript we carry out this task by considering a system as $CO_2^+$-IRMS. As "$^{S/W}\boldsymbol{\delta}_i \rightarrow {}^{S/D}\boldsymbol{\delta}_i$" conversion is a post-measurement-theoretical task, we here examine the designs of typical **δ**-scale-conversion processes, and clarify how accuracy of $^{S/D}\boldsymbol{\delta}_i$ can be ascertained on measured-data ($^{S/W}\boldsymbol{\delta}_i$ **…**). We present a model for a priori predicting output-uncertainty leading to the selection of a proper evaluation-method.

Our study shows that scale-conversion, even with the aid of *auxiliary*-reference-standard(s) ***Ai(s)***, cannot make "$^{S/D}\boldsymbol{\delta}_i$" free from ***W*** used for measurements; and the "$^{S/W}\boldsymbol{\delta}_i \xrightarrow{A1,A2} {}^{S/D}\boldsymbol{\delta}_i$" conversion-formula normally used in the literature is invalid. Besides, the latter-relation has been worked out, which leads to e.g., $\boldsymbol{f_J}$([$^{S/W}\boldsymbol{\delta}_J^{CO_2}$ ± *p*%],[$^{A1/W}\boldsymbol{\delta}_J^{CO_2}$ ± *p*%],[$^{A2/W}\boldsymbol{\delta}_J^{CO_2}$ ± *p*%]**)** = **(**$^{S/D}\boldsymbol{\delta}_J^{CO_2}$ ± **4.5***p*%**)**; whereas $\boldsymbol{F_J}$**(**[$^{S/W}\boldsymbol{\delta}_J^{CO_2}$ ± *p*%],[$^{A1/W}\boldsymbol{\delta}_J^{CO_2}$ ± *p*%]**)** = **(**$^{S/D}\boldsymbol{\delta}_J^{CO_2}$ ± **1.2***p*%**)**. That is, contrary to the general belief (*Nature* **1978**, *271*, 534; *Anal. Chem.* **2006**, *78*, 2439; *RCM*. **2007**, *21*, 3006), the scale-conversion by employing *one* than two ***Ai***-standards should ensure "$^{S/D}\boldsymbol{\delta}_i$" to be more accurate. However, a more valuable finding is that the transformation of any **δ**-estimate into its absolute value helps improve accuracy (viz.**:** [$^{S/D}\boldsymbol{\delta}_J^{CO_2}$ ± **1.2***p*%] → [$^{S}\boldsymbol{R}_J^{CO_2}$ ± **0.014***p*%]), or any reverse-process enhances uncertainty (e.g.: [$^{S}\boldsymbol{E}_{13/12}$ ± **0.015**$p$%] → [$^{S/D}\boldsymbol{\delta}_{13/12}$ ± **1.3**$p$%]). Thus, equally accurate though the absolute-estimates of isotopic-$CO_2$ and constituent-elemental-isotopic abundance-ratios ($\{^{S}\boldsymbol{R}_i^{CO_2}\}$ and $\{^{S}\boldsymbol{E}_d\}$, respectively) could be, in contradistinction any differential-estimate "$^{S/D}\boldsymbol{\delta}_i^{CO_2}$"or "$^{S/D}\boldsymbol{\delta}_d$" is shown to be *less* accurate. Further, for $\boldsymbol{S}$ and $\boldsymbol{D}$ to be similar, any absolute estimate is shown to turn out nearly absolute accurate but any "$^{S/D}\boldsymbol{\delta}$" value as really absurd. That is, estimated source specific absolute (e.g.: $^{S1}\boldsymbol{E}_d$, $^{S2}\boldsymbol{E}_d$ **…**) values, rather than corresponding differential results ($^{S1/D}\boldsymbol{\delta}_d$, $^{S2/D}\boldsymbol{\delta}_d$ **…**), should really represent their sources ($\boldsymbol{S1}$, $\boldsymbol{S2}$ **…**), and/ or be closely intercomparable.

## 1. INTRODUCTION

The natural isotopic abundance variations of lighter elements (which offer clues towards developing the fields as environmental, biochemical and earth sciences) are generally measured by the isotope ratio mass spectrometry (IRMS). The sensitivity and speed of such measurements involving complex matrices are revolutionized by the continuous-flow-IRMS.[1] Yet a measured data could be at an error, i.e. a measurement-error might get reflected as a source-variation or the vice-versa. Thus, for ensuring a variation-assessment to be authentic, any IRMS measurement is accomplished as a **ratio** of *similarly measured values* of abundance-ratios (viz. $^{S}\boldsymbol{R}_i$ and $^{W}\boldsymbol{R}_i$, of an isotopic pair "$\boldsymbol{i}$" in a sample-of-interest $\boldsymbol{S}$ and a working-lab-reference $\boldsymbol{W}$, respectively), and the result is expressed as a relative difference $^{S/W}\boldsymbol{\delta}_i$ (i.e. generally as: $^{S/W}\boldsymbol{\delta}_i$ = **(**[$^{S}\boldsymbol{R}_i$ / $^{W}\boldsymbol{R}_i$] **− 1)**). However, if only $\boldsymbol{W}$ is varied from lab to lab, the corresponding results cannot be inter-compared. This demands that a result should be reported with reference to (instead of $\boldsymbol{W}$) a recommended standard (say, $\boldsymbol{D}$).[2,3] That is the scale-conversion: $^{S/W}\boldsymbol{\delta}_i \rightarrow {}^{S/D}\boldsymbol{\delta}_i$ (where: $^{S/D}\boldsymbol{\delta}_i$ = **(**[$^{S}\boldsymbol{R}_i$ / $^{D}\boldsymbol{R}_{i,}$] **− 1)**) is an integral part of what is called as the IRMS. However, the involving of a computational-step in any measurement may turn out to mean the incorporation of an extraneous error source/ sink in the process.[4-6] We have, that is why, decided here to examine whether really the scale-conversion helps in enhancing accuracy and/ or comparability in results.

Further, the monitor isotopic species for measurements, and the isotopic species to actually be measured, may not be the same. For example, the isotopic analysis of carbon and/ or oxygen is used to be carried out indirectly as (the correspondingly generated) isotopic $CO_2^+$ ions.[7] Thus, before the shaping of a desired result as "$^{S/D}\boldsymbol{\delta}_d$", measured data on isotopic $CO_2^+$ ions ($^{S/W}\boldsymbol{\delta}_i^{CO_2}(s)$) are to be transformed (via scale-conversion: $^{S/W}\boldsymbol{\delta}_i^{CO_2}(s) \rightarrow {}^{S/D}\boldsymbol{\delta}_i^{CO_2}(s)$, and evaluation of absolute abundance-ratio(s) $^{S}\boldsymbol{R}_i^{CO_2}(s)$) into the constituent elemental isotopic abundance-ratio(s) $^{S}\boldsymbol{E}_d(s)$. Yet, the task: $^{S/W}\boldsymbol{\delta}_i^{CO_2} \rightarrow {}^{S/D}\boldsymbol{\delta}_i^{CO_2} \rightarrow {}^{S}\boldsymbol{R}_i^{CO_2} \rightarrow {}^{S}\boldsymbol{E}_d \rightarrow {}^{S/D}\boldsymbol{\delta}_d$ is a *theoretical* one, i.e. $^{S/D}\boldsymbol{\delta}_d$ should not vary for how exactly it is arrived at. However, the basic $CO_2^+$-IRMS evaluation: $^{S}\boldsymbol{R}_i^{CO_2} \rightarrow {}^{S}\boldsymbol{E}_d$ involves certain assumption.[7-14] Perhaps, that is why, $CO_2^+$-IRMS results are reported along with evaluation-parameters but, generally, without indicating the possible (i.e. measurement and assumption inflicted) variations. Moreover, it is often argued in the literature that scale conversion by employing two or more ***auxiliary***-reference-standards ($\boldsymbol{Ai}$, $\boldsymbol{i} = \mathbf{1}, \mathbf{2} \ldots$) "$^{S/W}\boldsymbol{\delta}_i \xrightarrow{\boldsymbol{A1},\boldsymbol{A2}\ldots} {}^{S/D}\boldsymbol{\delta}_i$" should ensure the result(s) to be accurate.[3,9,15,16,17] However, we feel it interesting to enquire why the involving of two additional *measured* data: ($^{S/W}\boldsymbol{\delta}_i$, $^{A1/W}\boldsymbol{\delta}_i$, $^{A2/W}\boldsymbol{\delta}_i$) $\rightarrow$ $^{S/D}\boldsymbol{\delta}_i$, should cause the $^{S/D}\boldsymbol{\delta}_i$-value to be more accurate than that evaluated by employing one or no $\boldsymbol{Ai}$. This work is therefore devoted towards examining the prevailing ideas in IRMS, and to offer a means for assessing possible variations in even stage specific $CO_2^+$-IRMS results, leading to the choice of a proper scale-conversion-method and/ or a design of evaluation (viz. whether "$^{S/D}\boldsymbol{\delta}_d$", or its absolute "$^{S}\boldsymbol{E}_d$", value should be the authentic tool for variation [based] studies).

Generally, evaluation means the *use of relevant relationship(s)* between measured (*independent*) and desired (*dependent*) variable(s) as ${}^{S/W}\boldsymbol{\delta}_i^{CO_2}\boldsymbol{(s)}$ and ${}^{S/D}\boldsymbol{\delta}_d\boldsymbol{(s)}$, respectively. Therefore, the foremost task below should be a consideration on the *formulae* involved in $CO_2^+$-IRMS. Then, we discuss certain typical evaluation-methods, and compare their features. However, the method of evaluation could be varied provided there is room for modifying "${}^{S/D}\boldsymbol{\delta}_d\boldsymbol{(s)}$ vs ${}^{S/W}\boldsymbol{\delta}_i^{CO_2}\boldsymbol{(s)}$" relationship(s). We thus also hint at the derivation of different method-governing relationships. Further, an output-estimate as ${}^{S/D}\boldsymbol{\delta}_d$ cannot be method-specific unless the achievable output-accuracy is controlled by the input-output relation(s) yielding ${}^{S/D}\boldsymbol{\delta}_d$.[4-6] Relevant evaluation methods are therefore studied below in terms of their *behaviors*.

It should however be noted that, for easy-distinguishing between *input* (i.e. **measured**) and *output* **δ**-variables, we will now re-denote "${}^{S/W}\boldsymbol{\delta}_i^{CO_2}$" as $\boldsymbol{X_i}$, and "${}^{S/D}\boldsymbol{\delta}_d$" as $\boldsymbol{Y_d}$; and refer to any output variable by "$\boldsymbol{Y}$". That is the *cascade of computational processes* (COCP): ${}^{S/W}\boldsymbol{\delta}_i^{CO_2} \rightarrow {}^{S/D}\boldsymbol{\delta}_i^{CO_2} \rightarrow {}^{S}\boldsymbol{R}_i^{CO_2} \rightarrow {}^{S}\boldsymbol{E}_d \rightarrow {}^{S/D}\boldsymbol{\delta}_d$, will be denoted as: $\boldsymbol{X_i} \rightarrow {}^{I}\boldsymbol{Y_d} \rightarrow {}^{II}\boldsymbol{Y_d} \rightarrow {}^{III}\boldsymbol{Y_d} \rightarrow \boldsymbol{Y_d}$. Further, output-error will be differentiated from measurement-error ($\Delta_i$) as "$\mathbf{Đ_d}$" (see Appendix **A**). We will also simplify the sample ($\boldsymbol{S}$) related notation "${}^{S}\boldsymbol{R}_i^{CO_2}$" as "$\boldsymbol{R_i}$", and "${}^{S}\boldsymbol{E_d}$" as "$\boldsymbol{E_d}$".

## 2. PRELIMINARIES

### 2.1 Uncertainty (accuracy)

Generally any measurement to be carried out is planned for, rather, eliminating error-sources. However, small *systematic* error-sources may not get even identified, and *accidental* errors can in no way be written off. Thus, ascertainment of true error (viz. $\Delta_i$, in any estimate, $\boldsymbol{x_i}$, of unknown $\boldsymbol{X_i}$) is impossible. However, if the **possible** variation in $\boldsymbol{x_i}$, i.e. clearly the *maximum* value (MV) of error $\Delta_i$, should not be known (indicated: $\boldsymbol{X_i} = [\boldsymbol{x_i} \pm |{}^{Max}\Delta_i|]$), then there will always be a doubt[18] as to whether "$x_i$" could [here, for the requirement as variation studies] be considered to represent "$X_i$". The MV of error (to be expected in an estimate $x_i$, and which should be established a priori in the process of choosing experimental methodology and by the help of relevant standards) is referred to as uncertainty [inaccuracy] or accuracy: $u_i$.[6]

Further, the extent of an error is signified only by its relative value.[19] So, we define: $\Delta_i = \frac{\Delta X_i}{X_i} = \frac{x_i - X_i}{X_i}$; and: $Đ_d = \frac{dY_d}{Y_d} = \frac{y_d - Y_d}{Y_d}$. Thus, by the measurement-uncertainty $u_i$, we mean: $u_i = |{}^{Max}\Delta_i|$. Similarly, output-uncertainty ($\varepsilon_d$) should equal to "$|{}^{Max}Đ_d|$". In any case, $u_i(s)$ should be established beforehand. Again, the relationship(s) of desired $Y_d(s)$ with measured $X_i(s)$ should be known. Therefore, uncertainty $\varepsilon_d$ (for a result $y_d$ to be obtained by a specified evaluation method) can be predicted,[6] i.e. an appropriate method of evaluation can be selected, a priori.

### 2.2 Basic relationships: (elemental vs molecular) isotopic abundance ratios

The relation of any measured (isotopic $CO_2$ abundance-ratio) $R_i$ with the desired (one or more of the constituent $^{13}C/^{12}C$, $^{17}O/^{16}O$ and $^{18}O/^{16}O$ abundance ratios) $E_1$, $E_2$ and $E_3$ (respectively), should be dictated, it may be pointed out, by the formula 'COO' itself:[20]

$$R_i = f_i(\{E_d\}),\ d = 1, 2, 3 \quad (1)$$

However, it can even theoretically[21] be seen that the $CO_2^+$ mass spectrum, for a limited (i.e. instrumentally achievable) resolution, consists of as many as six isotopic peaks (***m/z*** 44-49). That is there should be *five* basic $R_i$-equations. However, all isobaric $CO_2^+$ species [of a specified ***m/z***] can also be theoretically identified.[21] That is, any $R_i$-formula (Eq. 1) can easily be derived, and/ or is well defined,[20,22] e.g.: $R_{45/44} = (E_1 + 2E_2)$; $R_{46/44} = (E_2 \times (2E_1 + E_2) + 2E_3)$; etc.

### 2.3 Natural isotopic $CO_2$ abundance pattern and ratio ($R_i$) measurement

It may also be theoretically visualized that the isotopic-$CO_2^+$ ***abundance***, for *natural* and/ or *near-natural* isotopic-abundances of the constituents (C and O), is accounted by [instead of six] practically three peaks (***m/z*** 44-46).[21] That is, accurate ratio ($\boldsymbol{R_i}$) measurement should generally be possible for [instead of all five] *two* independent pairs only (viz.**:** $\boldsymbol{i}$ = $\boldsymbol{J}$ and $\boldsymbol{K}$, with either [$\boldsymbol{J}$ = 45/44 and $\boldsymbol{K}$ = 46/44] or [$\boldsymbol{J}$ = 45/44 and $\boldsymbol{K}$ = 46/45] or [$\boldsymbol{J}$ = 46/44 and $\boldsymbol{K}$ = 46/45]).

### 2.4 Basic evaluation process: $R_i(s) \rightarrow E_d(s)$, and sources of bias (if any)

Let us consider the evaluation of e.g. $^{13}C/^{12}C$ abundance ratio ($\boldsymbol{E_1}$) from an estimate ($\boldsymbol{r_{45/44}}$) of "$\boldsymbol{R_{45/44}}$". The evaluation**:** $\boldsymbol{E_1 = (r_{45/44} - 2E_2)}$, requires $\boldsymbol{E_2}$ (true $^{17}O/^{16}O$ abundance ratio) to be known. However (for an unknown sample), $\boldsymbol{E_2}$ cannot be known in advance. Therefore, the best means [for determining even single $\boldsymbol{E_d}$] should be to evaluate all three ratios ($\boldsymbol{E_1}$, $\boldsymbol{E_2}$ and $\boldsymbol{E_3}$**)** as the solutions of a set of equations**:** $\boldsymbol{f_i(\{E_d\}) = R_i}$ (with**:** $\boldsymbol{i, d}$ = 1, 2 and 3). Clearly, this process will require $\boldsymbol{R_i}$-measurements for three different isotopic-$CO_2$ pairs (**say:** $\boldsymbol{i}$ = $\boldsymbol{J}$, $\boldsymbol{K}$, and $\boldsymbol{L}$), but which also, as indicated above, cannot really be met. Thus the flawless ($CO_2^+$-IRMS) evaluation has had ever been difficult, and starting with Craig,[7] the problem is used to be resolved by employing (in place of the 3rd, i.e. $\boldsymbol{R_L}$ equation) an ad hoc relationship as Eq. 4 below:

$$f_J(E_1, E_2, E_3) = R_J \tag{2}$$

$$f_K(E_1, E_2, E_3) = R_K \tag{3}$$

$$E_2 = [{}^{D}E_2 / ({}^{D}E_3)^{\beta}]\,(E_3)^{\beta} \tag{4}$$

where $\boldsymbol{{}^{D}E_2}$ and $\boldsymbol{{}^{D}E_3}$, and $\boldsymbol{\beta}$ are known constants.

(Clearly, by the projected principle) no additional unknown is involved in Eq. 4, and is why it is possible to solve (*the set of*) Eqs. 2-4 for $\boldsymbol{\{E_d\}}$, $\boldsymbol{d =}$ 1, 2 and 3.

#### 2.4.1 *Can* "β" *really be a constant*?

$\boldsymbol{{}^{D}E_1}$ (not involved above), $\boldsymbol{{}^{D}E_2}$ and $\boldsymbol{{}^{D}E_3}$ (cf. Eq. 4) represent $^{13}C/^{12}C$, $^{17}O/^{16}O$ and $^{18}O/^{16}O$ abundance ratios, respectively, in the (recommended) standard $CO_2$ gas $\boldsymbol{D}$, and thus cannot be varied between labs. Therefore, for handling a given homogeneous sample material $\boldsymbol{S}$ (i.e. for **unknown** but *fixed* values of $\boldsymbol{E}_1$, $\boldsymbol{E}_2$ and $\boldsymbol{E}_3$), "β" should also, irrespective of lab, be fixed (cf. Eq. 4). Unfortunately, "$\boldsymbol{E}_2$ and $\boldsymbol{E}_3$" (i.e. their true values), and hence true "β", can never be known. Thus Eq. 4, although helps solve a genuine problem, is a *source of bias* in analysis. Moreover, a given β cannot be appropriate for all natural samples (which may differ from one another by source, i.e. by isotopic abundances). Besides, different researchers had recommended different values for "β" (e.g. 0.5,[7-10] 0.516,[11] 0.528,[12] etc.).

However, as already shown elsewhere,[22] the process of evaluation itself could be the best guide in setting β (and/ or Eq. 4), i.e. in arriving at (corresponding to any [given/ measured] estimates of $\boldsymbol{R}_J$ and $\boldsymbol{R}_K$) the *best* representative estimates of $\boldsymbol{E}_1$, $\boldsymbol{E}_2$ and $\boldsymbol{E}_3$. In other words, a possible means for avoiding β-specific biases in results is also known. Yet, β-value is usually chosen beforehand. Therefore, we will show below how the variations in results, due to any possible error in even the chosen "β", could be ascertained.

### 2.5 The $CO_2^+$-IRMS-evaluation scheme

As clarified by Eqs 2-4, the measurement should aim at estimating $\boldsymbol{R}_J$ and $\boldsymbol{R}_K$. However the IRMS variable is, as mentioned above, $\boldsymbol{X}_i$ (where: $\boldsymbol{X}_i = ([\boldsymbol{R}_i/{}^W\boldsymbol{R}_i] - 1)$, $\boldsymbol{i} = \boldsymbol{J}$ or $\boldsymbol{K}$).[8,9,15] Further, the desired variable is: $\boldsymbol{Y}_d$ (with: $\boldsymbol{Y}_d = ([\boldsymbol{E}_d/{}^D\boldsymbol{E}_d] - 1)$, $\boldsymbol{d} = \mathbf{1}$, or $\mathbf{2}$, or $\mathbf{3}$). Thus the different evaluation stages (COCP: $\boldsymbol{X}_i(s) \rightarrow\rightarrow\rightarrow\rightarrow \boldsymbol{Y}_d(s)$) involved in $CO_2^+$-IRMS are as follows:

**1st** *Stage* ($\boldsymbol{X}_i \rightarrow {}^I\boldsymbol{Y}_d$; with: ${}^I\boldsymbol{Y}_d = ([\boldsymbol{R}_i/{}^D\boldsymbol{R}_i] - 1)$, $\boldsymbol{i} = \boldsymbol{d} = \boldsymbol{J}, \boldsymbol{K}$): Any measured difference ($\boldsymbol{x}_i$) is required, i.e. by the principle called as IRMS, to be translated into the estimated difference (${}^I\boldsymbol{y}_d$) from the standard $\boldsymbol{D}$. However, ratios of isotopic-abundances can never be zero ($\boldsymbol{R}_i \neq \mathbf{0}$; ${}^D\boldsymbol{R}_i \neq \mathbf{0}$ and ${}^W\boldsymbol{R}_i \neq \mathbf{0}$). Thus "${}^I\boldsymbol{Y}_d$" can be expressed (and hence its value could be computed from $\boldsymbol{x}_i$) as:

$$ {}^I Y_d = \left( \frac{R_i}{{}^W R_i} \times \frac{{}^W R_i}{{}^D R_i} - 1 \right) = \left( [X_i + 1] \times \left[ {}^{W/D}C_i + 1 \right] - 1 \right) = {}^I f_d (X_i), d = i = J, K \tag{5} $$

where $^{D}\boldsymbol{R_i}$ is the $i^{th}$ isotopic $CO_2$ abundance ration in the gas ***D***, and**:** $^{W/D}\boldsymbol{C_i}$ = **(**[$^{W}\boldsymbol{R_i}$/$^{D}\boldsymbol{R_i}$] **– 1**).

Clearly, $^{W/D}\boldsymbol{C_i}$ should be known. That is the working reference ***W*** must be calibrated against the reference standard ***D***. However, this task could be avoided by making use of data **(**measured under identical possible experimental conditions as those employed for the sample ***S*****)** on one or more calibrated ***auxiliary*** reference standards (***A*1**, ***A*2**, etc).[8,9,15] In other words, Eq. 5 could also be replaced by a similar one which involves auxiliary measured variables as $\boldsymbol{Z1_i}$ (where**:** $\boldsymbol{Z1_i}$ = **([**$^{A1}\boldsymbol{R_i}$/$^{W}\boldsymbol{R_i}$**] – 1)); $\boldsymbol{Z2_i}$** (with: $\boldsymbol{Z2_i}$ = **([**$^{A2}\boldsymbol{R_i}$/$^{W}\boldsymbol{R_i}$**] – 1)**; etc.  However, for clarity, the details are returned to below.

**2nd** *Stage* ($^{I}\boldsymbol{Y_d} \rightarrow {}^{II}\boldsymbol{Y_d}$)**:** As the *required* ***data*** (cf. Eq. 2/ 3) is "$\boldsymbol{R_i}$", the **δ-**estimate ($^{I}\boldsymbol{y_d}$) should be translated into its *absolute* value ($\boldsymbol{r_i}$). Besides**:** $^{I}\boldsymbol{Y_d}$ **= ([**$\boldsymbol{R_i}$/$^{D}\boldsymbol{R_i}$**] – 1)**, and $^{D}\boldsymbol{R_i}$ is known, i.e. the retracing of $\boldsymbol{R_i}$ is a simple task**:** $\boldsymbol{R_i} = {}^{D}\boldsymbol{R_i} \times ({}^{I}\boldsymbol{Y_d} + 1)$. However, as the present tasks are required for simply translating the measured data "$\boldsymbol{x_i}$" into the desired data "$\boldsymbol{r_i}$", we refer to the 1st and 2nd stages as the **data-shaping** method, and re-denote the 2nd stage [output] variable "$\boldsymbol{R_i}$" as "$^{II}\boldsymbol{Y_d}$":

$$^{II}Y_d = {}^{II}f_d({}^{I}Y_d) = {}^{D}R_i \times ({}^{I}Y_d + 1), \quad d = i = J, K \tag{6}$$

Further, "$\boldsymbol{X_i} \xrightarrow{\boldsymbol{A1}} {}^{I}\boldsymbol{Y_d} \rightarrow {}^{II}\boldsymbol{Y_d}$" (i.e. the data-shaping by employing only one auxiliary reference standard ***A*1**) may be referred to as the **standardization**, and "$\boldsymbol{X_i} \xrightarrow{\boldsymbol{A1,A2}} {}^{I}\boldsymbol{Y_d} \rightarrow {}^{II}\boldsymbol{Y_d}$" (i.e. that which makes use of two auxiliary standards ***A*1** and ***A*2**) as the **normalization**.[9]

**3rd** *Stage* (**{**$^{II}\boldsymbol{Y_d}$**}** → **{**$^{III}\boldsymbol{Y_d}$**}):** Eqs. 2 and 3 are substituted for $\boldsymbol{R_i}$ by its estimate $^{II}\boldsymbol{y_d}$ (***i*** = ***d*** = ***J*** and ***K***), and then solved for the estimates of elemental isotopic abundance ratios (**{**$\boldsymbol{E_d}$**}**, ***d*** = 1-3) with the help of Eq. 4. Thus, Eqs 2-4 may (in terms of desired solutions) be rewritten as:

$$E_d = {}^{III}g_d(\{R_i\}) = {}^{III}g_d(\{{}^{II}Y_i\}), (d = 1, 2, 3), \text{ and } (i = J, K, L, \text{ with: } {}^{II}Y_L = R_L = \beta)$$

However, as $\boldsymbol{E_d}$ is the **3rd** stage (output) variable, it is re-denoted as $^{III}\boldsymbol{Y_d}$:

$$^{III}Y_d = {}^{III}g_d(\{R_i\}) = {}^{III}g_d(\{{}^{II}Y_i\}), (d = 1, 2, 3), \text{ and } (i = J, K, L, \text{ with: } {}^{II}Y_L = R_L = \beta) \tag{7}$$

$\mathbf{4^{th}}$ *Stage* ($^{III}\boldsymbol{Y_d} \rightarrow \boldsymbol{Y_d}$, *shaping of a desired result*)**:** The absolute estimate $^{III}\boldsymbol{y_d}$ is, for reporting, expressed as the relative difference ($\boldsymbol{y_d}$) from its recommended reference value $^{D}\boldsymbol{E_d}$:

$$Y_d = {}^{IV}f_d({}^{III}Y_d) = ([{}^{III}Y_d / {}^{D}E_i] - 1), \quad d = i = 1, 2, 3 \tag{8}$$

## 2.5.1 *Implications*

It can even independently (i.e. by treating a specific standard as the sample $\boldsymbol{S}$, and another as the gas $\boldsymbol{W}$) be verified that**:** $\boldsymbol{R_i} \equiv {}^{II}\boldsymbol{Y_d}$ ($\boldsymbol{i} = \boldsymbol{d} = \boldsymbol{J}, \boldsymbol{K}$; cf. Eqs. 2-3 and Eqs. 6-7), and**:** $\boldsymbol{E_d} \equiv {}^{III}\boldsymbol{Y_d}$ (cf. Eqs. 2-4 and Eqs. 7-8). That is, although the required sample *data* are designed to be shaped through processes as Eqs. 5-6, and the *results* as Eq. 8, the evaluation is really represented by a set of simultaneous equations as nos. **2-4** (or, in terms of desired solutions, by **no. 7**).

However the important point is that, by a measured estimate $\boldsymbol{x_i}$, it should be meant**:** $\boldsymbol{x_i} = (\boldsymbol{X_i} + \textbf{Error}) = (\boldsymbol{X_i} + \boldsymbol{\Delta_i})$, rather**:** $\boldsymbol{x_i} = (\boldsymbol{X_i} \pm \textbf{uncertainty}) = (\boldsymbol{X_i} \pm \boldsymbol{u_i})$. Therefore, a desired result $\boldsymbol{y_d}$ (which is obtained by the processes as Eqs. 5-8) would at best imply that**:** $\boldsymbol{y_d} = (\boldsymbol{Y_d} + \textbf{Đ}_\mathbf{d})$, and/ or**:** $\boldsymbol{y_d} = (\boldsymbol{Y_d} \pm \boldsymbol{\varepsilon_d})$; where $\textbf{Đ}_\mathbf{d}$ stands for true-error and $\boldsymbol{\varepsilon_d}$ for uncertainty in $\boldsymbol{y_d}$.

Further, the evaluation of $\{\boldsymbol{y_d}\}$ means the incorporation of *desired* **systematic** changes (i.e. those based on the *given* **relationships** of $\{\boldsymbol{Y_d}\}$ with $\{\boldsymbol{X_i}\}$, viz. Eqs. 5-8) in $\{\boldsymbol{x_i}\}$. Thus, result-specific true-errors $\{\textbf{Đ}_\mathbf{d}\}$ and uncertainties $\{\boldsymbol{\varepsilon_d}\}$ will also represent *desired* [e.g. Eqs. 5-8 dictated] variations in $\{\boldsymbol{\Delta_i}\}$ and $\{\boldsymbol{u_i}\}$, respectively;[4-6] and it is, therefore, intended to find out below whether $\boldsymbol{\varepsilon_d}$ equals to $\boldsymbol{u_i}$. In other words, the question raised is**:** what is the purpose that the processes like Eqs. 5-6 and Eq. 8 are designed to serve? Moreover, say that a lab [source] specific data-set ($\{\boldsymbol{x_i}\}^{LAB.1}$) differs from another data-set ($\{\boldsymbol{x_i}\}^{LAB.2}$) by 0.01%. Then should, at least for employing a given algorithm (as Eqs. 5-8), the variation between lab-results **(**$\{\boldsymbol{y_d}\}^{LAB.1}$ and $\{\boldsymbol{y_d}\}^{LAB.2}$**)** be 0.01%? Alternatively, can we a priori predict the measurement-accuracy $\boldsymbol{u_i}$ to be required for limiting the variations of "$\{\boldsymbol{y_d}\}^{LAB.1}$" from "$\{\boldsymbol{y_d}\}^{LAB.2}$" by 0.01%?

### 2.5.2 *Predictive means: parameter(s) characterizing a "$Y_d$ vs $X_i$(s)" relationship*

By given a relationship: $\boldsymbol{Y_d = f_d(\{X_i\}), i = 1, 2, \ldots N}$; it is meant that the relative rates ($\{\boldsymbol{M_i^d}\}$) of variation of $\boldsymbol{Y_d}$ as a function of $\{\boldsymbol{X_i}\}$ are also given:[6,20,22]

$$M_i^d = \left(\frac{\partial Y_d}{\partial X_i}\right)\left(\frac{X_i}{Y_d}\right) = \left(\frac{\partial Y_d/Y_d}{\partial X_i/X_i}\right), \quad i = J, K \ldots N \text{ (for a given } d) \tag{9}$$

Thus, how really the output uncertainty ($\boldsymbol{\varepsilon_d}$) is governed is also clarified:[6,22,23]

$$\varepsilon_d = \sum_{i=1}^{N} \left|M_i^d\right| u_i = \left(\sum_{i=1}^{N} \left|M_i^d\right| F_i\right) {}^{G}u = [UF]_d\ {}^{G}u \tag{10}$$

where $\boldsymbol{{}^{G}u}$ is a given $\boldsymbol{u_i}$-value **(**viz. which could be preset to be achieved before developing the required measurement-technique(s), i.e. before establishing actual methods- and/ or $\{\boldsymbol{X_i}\}$-specific $\{\boldsymbol{u_i}\}$**)**, so that**:** $\boldsymbol{F_i = (u_i / {}^{G}u)}$; and the ratio "$\boldsymbol{\varepsilon_d / {}^{G}u}$" is called as the uncertainty factor ($\boldsymbol{[UF]_d}$):[23]

$$[UF]_d = (\varepsilon_d / {}^{G}u) = \sum_{i=1}^{N} \left|M_i^d\right| F_i \tag{11}$$

If**:** $\boldsymbol{u_1 = u_2 \ldots = u_N (= {}^{G}u)}$, i.e. if**:** $\boldsymbol{F_i = 1 (i = 1, 2, \ldots N)}$, then**:** $\boldsymbol{[UF]_d = \sum_{i=1}^{N} \left|M_i^d\right|}$. However, the measurement-uncertainty $\boldsymbol{u_i}$ might vary as a function of $\boldsymbol{X_i}$. In that case, the factor "$\boldsymbol{F_i}$" can help to a priori assess "$\boldsymbol{[UF]_d}$", and hence to properly design all required experiments.

The desired result, $\boldsymbol{y_d}$, is obtained by the COCP as Eqs. 5-8. Therefore the requirement for "$\boldsymbol{y_d}$" to be **equally** accurate as the measured data "$\boldsymbol{x_i}$" is, as indicated by Eq. 10/ 11, that (for the COCP as a whole) "$\boldsymbol{[UF]_d} = \mathbf{1}$", i.e. the COCP should leave no signature of its involvement. If, however, "$\boldsymbol{[UF]_d}$" should be **<1**, then different lab-results ($\boldsymbol{y_d}^{\text{Lab1}}$, $\boldsymbol{y_d}^{\text{Lab2}}$ **…**) will be more accurate, and more closely intercomparable, than their lab-data ($\boldsymbol{x_i}^{\text{Lab1}}$, $\boldsymbol{x_i}^{\text{Lab2}}$ **…**).

## 3. SCALE-CONVERSION BY USING AN AUXILIARY STANDARD (*A*1)

The definition**:** $\boldsymbol{{}^{I}Y_d = ([R_i/{}^{D}R_i] - 1)}$ may, like Eq. 5, also include an auxiliary standard (***A*1**):

$${}^{I}Y_d = \left(\frac{R_i}{{}^{W}R_i} \times \frac{{}^{W}R_i}{{}^{A1}R_i} \times \frac{{}^{A1}R_i}{{}^{D}R_i}\right) - 1 = \left(\frac{\left(\left[R_i / {}^{W}R_i - 1\right] + 1\right) \times \left(\left[{}^{A1}R_i / {}^{D}R_i - 1\right] + 1\right)}{\left(\left[{}^{A1}R_i / {}^{W}R_i - 1\right] + 1\right)} - 1\right)$$

$$= \left( \frac{(X_i + 1) \times (C1_i + 1)}{(Z1_i + 1)} - 1 \right) = {}^{I}f_d(X_i, Z1_i), \quad d = i = J, K \qquad (5a)$$

where $\mathbf{C1_i}$ stands for a known isotopic calibration constant ($\mathbf{C1_i = ({}^{A1}\boldsymbol{R_i}/{}^{D}\boldsymbol{R_i}) - 1}$).

Again: $\boldsymbol{{}^{II}Y_d = f_d({}^{I}Y_d)}$, cf. Eq. 6. Thus, the standardization: $\boldsymbol{(X_i, Z1_i) \rightarrow {}^{I}Y_d \rightarrow {}^{II}Y_d}$ is usually carried out stage by stage as Eq. **5a** and Eq. **6**.[9,15] However, it really represents a single task:

$${}^{II}Y_d = {}^{II}f_d(X_i, Z1_i) = {}^{D}R_i \left( \frac{(X_i + 1) \times (C1_i + 1)}{(Z1_i + 1)} \right), \quad d = i = J, K \qquad (6a)$$

That is the desired results can truly be evaluated by a 3-stage-COCP (as Eqs. **6a**, **7** and **8**). Similarly the results remain unaltered whether all relative differences are expressed in terms of either *unity* as here,[9] or *percentage*, or per *mil*[2,3,7,8,15]. Further, if the auxiliary variable is defined as: $\boldsymbol{{}^{W/A1}Z1_i = ([{}^{W}R_i/{}^{A1}R_i] - 1)}$,[8] then the form of Eq. 6a (but ***not*** the output variable, $\boldsymbol{{}^{II}Y_d}$, itself) will be different: $\boldsymbol{{}^{II}Y_d = {}^{II}f_d(X_i, {}^{W/A1}Z1_i) = ({}^{D}R_i \times [X_i + 1] \times [{}^{W/A1}Z1_i + 1] \times [C1_i + 1])}$.

It should however be noted that we here consider: $\boldsymbol{J = 45/44}$, and $\boldsymbol{K = 46/44}$, i.e. the basic set of equations (no. **2-4**) to be as:

$${}^{III}Y_1 + 2\,{}^{III}Y_2 = {}^{II}Y_J = {}^{II}Y_{45/44} \qquad (2a)$$

$${}^{III}Y_2\,(2\,{}^{III}Y_1 + {}^{III}Y_2) + 2\,{}^{III}Y_3 = {}^{II}Y_K = {}^{II}Y_{46/44} \qquad (3a)$$

$${}^{III}Y_2 = [{}^{D}E_2/({}^{D}E_3)^{\beta}]\,({}^{III}Y_3)^{\beta} \qquad (4a)$$

### 3.1 Comparability: roles of data (Eq. 6a) and result-shaping (Eq. 8) processes

#### 3.1.1 *Experimental viewpoint*

We consider the case by Verkouteren and Lee,[9] i.e. assume the true $\boldsymbol{X_i}$ and $\boldsymbol{Z1_i}$ values to equal their respective measured estimates[9] ($\boldsymbol{X_J = x_J}$ = –0.010550, $\boldsymbol{Z1_J = z1_J}$ = –0.003220, $\boldsymbol{X_K}$ = $\boldsymbol{x_K}$ = –0.011820, and $\boldsymbol{Z1_K = z1_K}$ = –0.008980) and the constants to also be the reported[9] those ($\boldsymbol{C1_J}$ = –0.004112, $\boldsymbol{{}^{D}R_J}$ = 11.99493320x$10^{-3}$, $\boldsymbol{C1_K}$ = –0.018499, $\boldsymbol{{}^{D}R_K}$ = 41.42979699x$10^{-4}$, $\boldsymbol{{}^{D}E_1}$ =

11.2372x10$^{-3}$, $^{D}\boldsymbol{E}_{2}$ = 37.8866601x10$^{-5}$, $^{D}\boldsymbol{E}_{3}$ = 20.67160680x10$^{-4}$, and $\beta$ = 0.5**)**. We here evaluate the desired results by using, instead of **Eqs. (5a**, **6**, **7** and **8)**,[9] **Eqs. (6a**, **7** and **8**); and present them (cf. example no. 00) in Table 1. However, our results are no different from the reported[9] estimates (example no. 0). This clarifies that, if the scale conversion method is chosen and if the required data are fixed, then the results cannot vary whether a 3- or 4-steps COCP is used.

3.1.1.1 Input to output variations

How the results (for **given:** a homogeneous material ***S***, achievable measurement-accuracy $^{G}\boldsymbol{u}$ and a data processing method) may *vary* as a function of measurement-*lab* and/ or-*time* is also exemplified in Table 1 (cf. nos. 1-5). It may however be pointed out that, for ascertaining the **extreme** variations to be expected in results, all the data ($\boldsymbol{x_i}$ and $\boldsymbol{z1_i}$ [$\boldsymbol{i = J}$ and $\boldsymbol{K}$], and even "$\boldsymbol{\beta}$") are varied by exactly $\pm^{G}\boldsymbol{u}$, with e.g. ($|\Delta_i|$ = $|^{Max}\Delta_i|$ =) $^{G}\boldsymbol{u}$ = **1**%. That is the variation between data (viz.**:** $\boldsymbol{x_J}^{\mathbf{Exp.1}}$ and $\boldsymbol{x_J}^{\mathbf{Exp.2}}$ in Table 1, or) from two different labs, which yield to a required accuracy $^{G}\boldsymbol{u}$, could be even "**2**$^{G}\boldsymbol{u}$". Clearly, if the measurement accuracy is varied between labs, "**2**$^{G}\boldsymbol{u}$" will equal to "$^{G}\boldsymbol{u}^{\mathbf{Lab1}} - (-^{G}\boldsymbol{u}^{\mathbf{Lab2}})$". In other words, the *measurement-comparability*, i.e. the **highest** "lab to lab variation ($|^{Max}\Delta_i^{Lab1}|$ + $|^{Max}\Delta_i^{Lab2}|$)" to be possible in a data as $\boldsymbol{x_i}$, is**:** **2**$^{G}\boldsymbol{u}$ (= 2%, cf. Table 1). Yet, it may be noted e.g. that**:** $\boldsymbol{y_3}^{\mathbf{Exp.1}}$ differs from $\boldsymbol{y_3}^{\mathbf{Exp.2}}$ by <**2**$^{G}\boldsymbol{u}$, but**:** $\boldsymbol{y_1}^{\mathbf{Exp.1}}$ varies from $\boldsymbol{y_1}^{\mathbf{Exp.2}}$ by >**2**$^{G}\boldsymbol{u}$; and/ or $\boldsymbol{y_2}^{\mathbf{Exp.3}}$ and $\boldsymbol{y_2}^{\mathbf{Exp.4}}$ are as different as ≈**4**$^{G}\boldsymbol{u}$. Thus, as graphically indicated in Fig. 1***a***, achievable *comparability-in-lab-results* may**:** **(i)** differ from measurement-comparability and**:** **(ii)** even vary as a function of desired output-variable.

3.1.1.2 Can a **δ**-estimate be more significant than its absolute value?

We first consider the **δ**-inputs and corresponding *absolute*-output of Eq. **6a** ($\boldsymbol{X_i} \xrightarrow{\boldsymbol{Z1_i}} {}^{II}\mathbf{Y}_i$; $\boldsymbol{i} = \boldsymbol{J}$ or $\boldsymbol{K}$). The net input error ($|^{X}\Delta_i|+|^{Z1}\Delta_i|$) is [as exemplified, cf. any of nos.**:** 1-5 in Table 1]**:** **2**$^{G}\boldsymbol{u}$ = 2%. Yet, the output-error "$|^{II}\mathbf{Đ}_J|$" is restricted as ≤0.014%, and "$|^{II}\mathbf{Đ}_K|$" is ≤0.021%;

i.e. the conversion of a **δ**-estimate (***x**$_i$*) into its absolute value ($^{II}y_i$) is observed to help improve accuracy. Moreover, the desired absolute *results* [i.e. outputs, **{**$^{III}y_d$**}**, of Eq. 7] have also turned out as accurate as the isotopic-$CO_2$ *data* ($^{II}y_J$ and $^{II}y_K$); e.g. error**:** $|^{III}Đ_3|$ = $|^{II}Đ_K|$. However any **δ**-result ($y_d$) is rather erroneous, i.e. "*absolute* to **δ**" conversion as Eq. **8** [e.g.**:** $^{III}y_1 \rightarrow y_1$] is shown to be accompanied by error-enhancement [($|^{III}Đ_1|$ = **0.015**%) → ($|Đ_1|$ = **1.3**%)]. Thus, though $^{III}y_1^{Exp.1}$ differs from $^{III}y_1^{Exp.2}$ by 0.030%, $y_1^{Exp.1}$ and $y_1^{Exp.2}$ are as different as 2.6%. That is, the absolute result $^{III}y_d$ **(**rather than its **δ-**estimate $y_d$**)** is signified here to *better* represent the source ***S***, and even to *more* closely compare with other such results (see also Fig. 1***a***).

Let's now say that the results as example no. 1 in Table 1 stand for an unknown source ***S*1**, and no. 2 for another source ***S*2**. Then the sources ***S*1** and ***S*2** should be [for judging by either the (absolute) $^{III}Y_d$- or the (differential) $Y_d$-values] *different***: (i)** from a source as the recommended standard ***D*** (which is represented by e.g.**:** $^{D}[^{III}Y_1] \equiv {}^{D}E_1$ = 0.0112372, and**:** $^{D}[Y_1] = {}^{D/D}\delta_1$ = **0.0**), and**: (ii)** at least apparently, from one another. Further, let us <u>*assume*</u> that the absolute results (e.g. $^{III}y_1^{Exp.1}$ and $^{III}y_1^{Exp.2}$, i.e. which differ from one another by **0.03**%) signify that both ***S*1** and ***S*2** are of *similar* geochemical origins. <u>*Then*</u> can we, even in terms of the differential results $y_1^{Exp.1}$ and $y_1^{Exp.2}$ [i.e. for tens of fold ***higher*** a variation**: 2.6**%], imply that both ***S*1** and ***S*2** bear similar histories? That is, shouldn't the differential results lead to rather wrong conclusions?

### 3.1.2 *Are the processes as Eq. 6a and Eq. 8 really opposite in nature*?

#### 3.1.2.1 <u>Process-specific uncertainty formulae</u>

We may, in terms of estimates, rewrite Eq. 6a as**:** $[^{II}y_d \pm {}^{II}\varepsilon_d] = {}^{II}f_d([x_i \pm {}^{X}u_i],[z1_i \pm {}^{Z1}u_i])$, with $^{II}\varepsilon_d$ as the 1st cumulative (i.e. 2nd stage) uncertainty, and $^{X}u_i$ and $^{Z1}u_i$ as $X_i$ and $Z1_i$ *measurement-uncertainties*, respectively. Further, as indicated by Eq. 10, $^{II}\varepsilon_d$ should be decided as:

$$^{II}\varepsilon_d = (|{}^{X}M_i^d| \times {}^{X}u_i) + (|{}^{Z1}M_i^d| \times {}^{Z1}u_i) =$$

$$(\left|{}^{X}M_i^d\right| \times {}^{X}F_i) + (\left|{}^{Z1}M_i^d\right| \times {}^{Z1}F_i)\ {}^{G}u = {}^{II}[UF]_d\ {}^{G}u, \qquad (d = i = J, K) \qquad (12)$$

where ${}^{X}\boldsymbol{M}_{\boldsymbol{i}}^{\boldsymbol{d}}$ and ${}^{Z1}\boldsymbol{M}_{\boldsymbol{i}}^{\boldsymbol{d}}$ are theoretical constants (e.g.**:** ${}^{X}M_i^d = \left(\partial\, {}^{II}Y_d / \partial X_i\right) \times \left(X_i / {}^{II}Y_d\right)$, ***d* = *i* = *J, K***; cf. Eq. 9) characterizing Eq. 6a; ${}^{X}F_i = ({}^{X}u_i / {}^{G}u)$ and ${}^{Z1}F_i = ({}^{Z1}u_i / {}^{G}u)$; and ${}^{II}[UF]_d$ is called as the 2nd stage (or 1st cumulative) uncertainty factor, and can be pre-evaluated as (cf. Eq. 11):

$${}^{II}[UF]_d = ({}^{II}\varepsilon_d / {}^{G}u) = (\left|{}^{X}M_i^d\right| \times {}^{X}F_i) + (\left|{}^{Z1}M_i^d\right| \times {}^{Z1}F_i), \qquad (d = i = J, K) \qquad (13)$$

However the outputs of Eq. 6a, i.e. "$({}^{II}y_J \pm {}^{II}\varepsilon_J)$ and $({}^{II}y_K \pm {}^{II}\varepsilon_K)$", and β (rather**:** $({}^{II}y_L \pm u_L) \equiv (\beta \pm u_\beta)$), constitute the inputs for the 3rd stage (i.e. for Eqs. 2a, 3a and 4a, respectively). Again, the 3rd stage outputs (${}^{III}y_d \pm {}^{III}\varepsilon_d$, ***d* = 1, 2, 3**) may always be expressed as **Eq. 7**. Thus the 3rd stage (i.e. 2nd cumulative) uncertainty ${}^{III}\varepsilon_d$ would also be governed as Eq. 10:

$${}^{III}\varepsilon_d = (\left|{}^{III}M_J^d\right| \times {}^{II}\varepsilon_J) + (\left|{}^{III}M_K^d\right| \times {}^{II}\varepsilon_K) + (\left|{}^{III}M_\beta^d\right| \times u_\beta) = ((\left|{}^{III}M_J^d\right| \times {}^{II}[UF]_J) +$$

$$(\left|{}^{III}M_K^d\right| \times {}^{II}[UF]_K) + (\left|{}^{III}M_\beta^d\right| \times F_\beta))\ {}^{G}u = {}^{III}[UF]_d\ {}^{G}u, \quad (d = 1, 2, 3) \qquad (12a)$$

where**:** ${}^{III}\boldsymbol{M}_{\boldsymbol{J}}^{\boldsymbol{d}}$, ${}^{III}\boldsymbol{M}_{\boldsymbol{K}}^{\boldsymbol{d}}$, and ${}^{III}\boldsymbol{M}_{\boldsymbol{\beta}}^{\boldsymbol{d}}$ **(**the evaluation of which is already discussed elsewhere[23]**)** stand for the predicted (cf. Eq. 9) rates-of-variations of ${}^{III}Y_d$ with ${}^{II}Y_J$, ${}^{II}Y_K$, and β, respectively; $F_\beta = (u_\beta / {}^{G}u)$; and ${}^{III}[UF]_d$ is the 2nd cumulative (3rd stage) uncertainty factor:

$${}^{III}[UF]_d = ({}^{III}\varepsilon_d/{}^{G}u) = (\left|{}^{III}M_J^d\right|\ {}^{II}[UF]_J) + (\left|{}^{III}M_K^d\right|\ {}^{II}[UF]_K) + (\left|{}^{III}M_\beta^d\right| F_\beta), \ (d = 1, 2, 3) \quad (13a)$$

If β should be treated as a constant, then**:** ${}^{III}\boldsymbol{M}_{\boldsymbol{\beta}}^{\boldsymbol{d}} = 0$. Thus, Eq. 13a should be rewritten as:

$${}^{III}[UF]_d = ({}^{III}\varepsilon_d/{}^{G}u) = (\left|{}^{III}M_J^d\right|\ {}^{II}[UF]_J) + (\left|{}^{III}M_K^d\right|\ {}^{II}[UF]_K), \quad (d = 1, 2, 3) \qquad (13a^{/})$$

The desired $Y_d$-value is, however, obtained at the end of Eq. 8**:** $(y_d \pm \varepsilon_d) = {}^{IV}f_d([{}^{III}y_d \pm {}^{III}\varepsilon_d]) = ([({}^{III}y_d \pm {}^{III}\varepsilon_d) / {}^{D}E_i] - 1)$. Therefore, the **COCP-uncertainty** ($\varepsilon_d$) should be decided as:

$$\varepsilon_d = (\left|M_d^d\right| \times {}^{III}\varepsilon_d) = (\left|M_d^d\right| \times {}^{III}[UF]_d \times {}^{G}u) = [UF]_d\ {}^{G}u, \qquad (d = 1, 2, 3) \qquad (12b)$$

where $\boldsymbol{M}_{\boldsymbol{d}}^{\boldsymbol{d}}$ is a constant representing the rate-of-variation (cf. Eq.9) of $Y_d$ as a function of ${}^{III}Y_d$, and $[UF]_d$ may be referred to as the **COCP** (i.e. 3rd cumulative/ 4th stage) **uncertainty factor**:

$$[UF]_d = (\varepsilon_d / {}^{G}u) = \left|M_d^d\right| \times {}^{III}[UF]_d \qquad (13b)$$

Further, considering "*uncertainty*" as the only variable, Eq. 12b might be rewritten as:

$$\varepsilon_d = f_d({}^{III}\varepsilon_d) = f_d(g_d({}^{II}\varepsilon_J, {}^{II}\varepsilon_K, u_\beta)) = f_d(g_d({}^{II}f_J({}^{X}u_J, {}^{Z1}u_J), {}^{II}f_K({}^{X}u_K, {}^{Z1}u_K), u_\beta)), \quad d = 1, 2, 3 \qquad (12b^{/})$$

Eq. 12b$^{/}$ clarifies that the uncertainty $\boldsymbol{\varepsilon_d}$ [of a desired **δ**-result, $y_d$] should represent a ***net*** of all respective stage specific ***systematic*** *changes* in the uncertainties**:** $^{X}u_J$, $^{Z1}u_J$, $^{X}u_K$, $^{Z1}u_K$ and $u_\beta$. For example, whether $y_d$ should be better representative than its absolute value $^{III}y_d$ is decided by "$f_d$" (cf. Eq. 8), i.e. by the $M_d^d$-value only. However**:** "$({}^{II}f_J, {}^{II}f_K) \rightarrow g_d \rightarrow f_d$", i.e. "$[UF]_d$" will govern whether $y_d$ should be more accurate than a measured data $x_i$.

Further, let us consider a case as that**:** $^{X}u_J = {}^{Z1}u_J = {}^{X}u_K = {}^{Z1}u_K = u_\beta = {}^{G}u$, i.e. $\{F_i = 1\}$. Then Eq. 13b should be rewritten as:

$$[UF]_d = \left(\left|M_d^d\right| \times {}^{III}[UF]_d\right) = \left(\left|M_d^d\right| \times \left(\left|{}^{III}M_J^d\right| \times \left[\left|{}^{X}M_J^d\right| + \left|{}^{Z1}M_J^d\right|\right] + \left|{}^{III}M_K^d\right| \times \left[\left|{}^{X}M_K^d\right| + \left|{}^{Z1}M_K^d\right|\right] + \left|{}^{III}M_\beta^d\right|\right)\right), \quad (d = 1, 2, 3) \qquad (13b^{/})$$

Clearly, "$[UF]_d$" should represent the **collective** *rate of variation* of the COCP as Eqs. (6a, 7 and 8), i.e. $\boldsymbol{\varepsilon_d}$ (cf. Eq. 12b) should be governed by rather the *nature* of the COCP *as a whole*.

3.1.2.2 Natures of the processes as Eqs. (6a, 7 and 8)**:** are the findings of Table 1 authentic?

The above discussed stage/ COCP specific parameters are, in terms of their governing factors and also (in terms of the system[9] as Table 1) numerically, presented in Table 2. However, if**:** $^{G}u$ = 1% (cf. Table 1), then the uncertainty "$\boldsymbol{\varepsilon_d}$" should numerically equal the uncertainty-factor "$[UF]_d$", e.g.**:** $\varepsilon_3 = ([UF]_3 \times {}^{G}u) = [UF]_3$ = 0.97% (cf. 3rd Cumulative Stage in Table 2 for $Y_3$). This thus supplements the indication (Table 1) that, if really all different $CO_2^+$ data in ref. [9] are equally accurate, then the estimated *differential* $^{18}O/^{16}O$ ratio ($y_3$) should be somewhat **more** accurate than a *measured* [differential isotopic-$CO_2^+$ abundance] ratio as $x_i$. Similarly [i.e. as implied by Table 1, viz.**:** $\left|{}^{Max}Đ_1\right|$ = 1.3% and**:** $\left|{}^{Max}Đ_2\right|$ = 1.97%], the *differential* $^{13}C/^{12}C$ and

$^{17}O/^{16}O$ ratios ($\boldsymbol{y_1}$ and $\boldsymbol{y_2}$) are asserted to be more ***inaccurate*** ($\boldsymbol{\varepsilon_1}$ = 1.4%; and**:** $\boldsymbol{\varepsilon_2}$ = 1.97%, respectively, cf. Table 2) than a measured ratio as $\boldsymbol{x_i}$. Besides, the small discrepancy as**:** $\boldsymbol{\varepsilon_1}$ = 1.4% (cf. Table 2) but**:** $|{}^{Max}\mathbf{Đ}_1|$ = 1.3% (cf. [Table 1) should be explicable because Eqs. 6a-8 are *non-linear* and the errors-in-data are *significant* (±1.0%).[6] Further, the uncertainty $\boldsymbol{\varepsilon_2}$ is ensured to be >$\boldsymbol{\varepsilon_3}$ (cf. Table 2). Yet, as shown by the example no. 1 or 2 in Table 1, error "$|\mathbf{Đ}_2|$" is "<$|\mathbf{Đ}_3|$". This rather signifies that any true error [viz. the *output*-error $\mathbf{Đ}_2$] can even [corresponding to **non-zero** *input* (measurement) errors ($\{|\Delta_i| \neq \mathbf{0}\}_i$)] turn out to be zero.[6]

Moreover, as shown, the rate-of-variation ($\boldsymbol{M_d^d}$) of any *differential* ratio, $\boldsymbol{Y_d}$, as a function of its *absolute* value, $\boldsymbol{{}^{III}Y_d}$, is governed as**:** $\boldsymbol{M_d^d = ({}^{III}Y_d / [{}^{III}Y_d - {}^{D}E_d])}$, i.e.**:** $|\boldsymbol{M_d^d}|$ is >**1** (cf. Table 2 for the 3rd cumulative stage). This explains why should the transformation as Eq. **8** (viz.**:** $\boldsymbol{{}^{III}y_3} \rightarrow \boldsymbol{y_3}$) be accompanied by error-**enhancement**. The predicted enhancement-factor (e.g.**:** [$\boldsymbol{\varepsilon_3 / {}^{III}\varepsilon_3}$] = [**0.97 / 0.021**] ≈ 46, cf. Table 2), and its observed value ([$|\mathbf{Đ}_3| / |{}^{III}\mathbf{Đ}_3|$] ≈ 46, cf. Table 1), are even identical. Further, Eq. **6a** is confirmed to be an error-***reducing*** process, i.e. why did any evaluated isotopic-$CO_2$ abundance-ratio, e.g. $\boldsymbol{{}^{II}y_K}$, turn out less erroneous ($|{}^{II}\mathbf{Đ}_K| \leq$ **0.021**%, cf. Table 1) than its differential estimate $\boldsymbol{x_K}$, is explained**:** ($\boldsymbol{{}^{II}\varepsilon_K / {}^{G}u}$) = **0.021**. Again, "$\boldsymbol{{}^{III}Y_d}$" is shown to be rather insensitive towards "β", i.e.**:** $|\boldsymbol{{}^{III}M_\beta^d}| \ll$ **1**; and/ or**:** $\boldsymbol{{}^{III}\varepsilon_d \approx {}^{II}\varepsilon_i}$ (cf. Table 2 for Eq. **7**). Thus, why should even the elemental isotopic abundance ratio, $\boldsymbol{{}^{III}y_d}$, turn out as accurate as any isotopic-$CO_2$ abundance ratio, $\boldsymbol{{}^{II}y_i}$ [viz. why**:** $|{}^{III}\mathbf{Đ}_3| = |{}^{II}\mathbf{Đ}_K|$; cf. Table 1], is explicable.

3.1.2.3 Does the scale-conversion as Eq. 5a itself help enhance accuracy?

The **1st** stage (Eq. 5a**:** $\boldsymbol{(X_i, Z1_i) \rightarrow {}^{I}Y_d}$; $\boldsymbol{i = d = J}$ and $\boldsymbol{K}$) and the **2nd** stage (Eq. 6**:** $\boldsymbol{{}^{I}Y_d \rightarrow {}^{II}Y_d}$) tasks are here collectively carried out by Eq. 6a. Yet, it could be shown that**: (i)** the **1st** stage uncertainties to be**:** $\boldsymbol{{}^{I}\varepsilon_J}$ = 1.2$\boldsymbol{{}^{G}u}$, and**:** $\boldsymbol{{}^{I}\varepsilon_K}$ = 0.962$\boldsymbol{{}^{G}u}$; and **(ii)** the 2nd stage process is characterized by a variation-rate ($|{}^{II}M^{d}_{d}|$) of $\ll$**1**. Actually**:** $^{II}M^{d}_{d}$ **= (**$^{I}Y_d$**/[**$^{I}Y_d$ **+ 1])**, which means that**:** $^{II}M^{J}_{J}$ **=** –0.0116, and $^{II}M^{K}_{K}$ = –0.0218, i.e. in turn that (**cf.** Eq. 10)**:** $^{II}\varepsilon_J$ = ($|{}^{II}M^{J}_{J}|$ × $^{I}\varepsilon_J$) = 0.014$^{G}u$, and**:** $^{II}\varepsilon_K$ = ($|{}^{II}M^{K}_{K}|$ × $^{I}\varepsilon_K$) = 0.021$^{G}u$. Thus, although the scale conversion as Eq. **5a** does not on its own cause a significant change in accuracy**:** ($^{I}\varepsilon_d/^{G}u$) ≈ **1** (see also Fig. **1*a***), the translation of a differential estimate ($^{I}y_d \pm {}^{I}\varepsilon_d$) into its absolute value ($^{II}y_d \pm {}^{II}\varepsilon_d$) as Eq. **6** should be accompanied by reduction (here, 40-90 fold) of uncertainty**:** ($^{II}\varepsilon_d/^{I}\varepsilon_d$) $\ll$ **1**, and/ or**:** ($^{II}\varepsilon_d/^{G}u$) $\ll$ **1**.

3.1.2.4 Can any estimate ever be independent of corresponding measurement reference "***W***"?

The *rate-of-variation* of "$^{II}Y_d$" (cf. Eq. 6a) as a function of a measured variable $X_i$ (or $Z1_i$) is shown (Table 2) to be decided by $X_i$ (or $Z1_i$) itself, i.e.**:** $^{X}M^{d}_{i}$ **=** $f_M(X_i)$; and/ or**:** $^{Z1}M^{d}_{i}$ **=** $f_M(Z1_i)$. However, for given the $CO_2$ gases ***S*** and ***A*1** to be measured, (true) $X_i$ and $Z1_i$ values will be fixed by the working reference gas ***W*** to be used (e.g.**:** $X_i$ = **(**[$R_i/^{W}R_i$] **– 1)**). Thus, though ***W*** is not required to be a standard, *achievable* accuracy ($^{II}\varepsilon_d$, and in turn $^{III}\varepsilon_d$, and then $\varepsilon_d$, cf. Table 2) of any desired **δ**-result as $y_d$ [i.e. **any** estimate**:** $^{II}y_d$, or $^{III}y_d$, or $y_d$ itself] will ***vary*** for varying the isotopic composition of the gas as (either ***S*** or ***A*1** or) even alone ***W***, however.

## 4. NORMALIZATION BY USING TWO AUXILIARY STANDARDS (*A*1 and *A*2)

### 4.1 The conventional normalization method

The normalization formula being used **(**cf. Eq. 19 in ref. [9], or Eq. 17 in [15]**)** is:

$$^{nI}Y_d = {}^{I}f_d\left(X_i, Z1_i, Z2_i\right) = \left(X_i - Z1_i\right) \times \left(\frac{C2_i - C1_i}{Z2_i - Z1_i}\right) + C1_i, d = i = J, K \quad (14)$$

where $^{nI}Y_d$ is referred to as the normalized output variable, $Z2_i$ **= ((**$^{A2}R_i/^{W}R_i$**) – 1**), and $C2_i$ stands (like $C1_i$ in Eq. 5a/ 6a) for isotopic calibration constant**:** $C2_i$ = **((**$^{A2}R_i/^{D}R_i$**) – 1**). However, as "$^{nI}Y_d$" is different from the *scale converted differential* ratio $^{I}Y_d$ (cf Eq. 5/ 5a), the 2nd stage variable (i.e. the *absolute* ratio, say, $^{nII}Y_d$) will also be different from $^{II}Y_d$ (cf. Eq. 6/ 6a):

$$^{nII}Y_d = {}^{II}f_d({}^{nI}Y_d) = {}^{D}R_i\,({}^{nI}Y_d + 1), \quad d = i = J, K \tag{15}$$

Or, in the cumulative sense (**[$X_i$, $Z1_i$, $Z2_i$] → $^{nII}Y_d$**):

$$^{nII}Y_d = {}^{II}f_d\left(X_i, Z1_i, Z2_i\right) = {}^{D}R_i \times \left[\left\{\left(X_i - Z1_i\right)\times\left(\frac{C2_i - C1_i}{Z2_i - Z1_i}\right) + C1_i\right\} + 1\right], d = i = J, K \tag{15a}$$

And, the 3rd stage (Eq. 7) and 4th stage (Eq. 8) processes may here be re-represented as:

$$^{nIII}Y_d = g_d(\{^{nII}Y_i\}), (d = 1, 2 \text{ and } 3), \text{ and } (i = J, K \text{ and } L, \text{ with: } {}^{nII}Y_L = \alpha) \tag{16}$$

$$^{n}Y_d = f_d({}^{nIII}Y_d) = [({}^{nIII}Y_d / {}^{D}E_i) - 1], \quad d = i = 1, 2, 3 \tag{17}$$

### 4.1.1 *Should Eq. 14 yield (more) accurate results (than by Eq. 5a/ 6a)?*

We, for a proper comparison between method specific results, consider the 2nd set of auxiliary variables (**{$Z2_i$}**) and constants (**{$C2_i$}**) to also be those reported by Verkouteren and Lee; [9] i.e.**:** $\boldsymbol{Z2_J = z2_J}$ **=** –0.028810, $\boldsymbol{C2_J}$ **=** –0.039998; $\boldsymbol{Z2_K = z2_K}$ **=** –0.024100 and $\boldsymbol{C2_K}$ **=** –0.033675. We present our evaluated results, for**: (i)** zero [cf. example no. 00], and**: (ii)** ±1% errors in inputs **[**i.e. like Table 1, for**:** $^{G}\boldsymbol{u}$ = **1**%, cf. example nos. 1-5**]**, in Table **3**.

However, any output in Table 3 is generally at a larger error than that in Table 1. For example, the ratio "$^{nII}\boldsymbol{y_J}$" is as varied as ±0.03%, but the variation of the ratio "$^{II}\boldsymbol{y_J}$" is restricted to ±0.014% (cf. Eq. **6a** and example nos. 1-5 in Table **1**). Similarly, the error **"**$^{Max}|{}^{nII}Đ_K|$ = 0.0245%" is higher than the error "$^{Max}|{}^{II}Đ_K|$ = 0.021%". Besides the COCP (i.e. Eqs. 14, 15, 16 and 17) specific parameters, which are tabulated in Table 4, confirm that the present method is not a better one; e.g.**:** $^{nII}\boldsymbol{\varepsilon_J}$ **=** 0.03$^{G}\boldsymbol{u}$, and**:** $^{nII}\boldsymbol{\varepsilon_K}$ **=** 0.024$^{G}\boldsymbol{u}$ (i.e. even though [cf. Table 2]**:** $^{II}\boldsymbol{\varepsilon_J}$ **=** 0.014$^{G}\boldsymbol{u}$, and**:** $^{II}\boldsymbol{\varepsilon_K}$ **=** 0.021$^{G}\boldsymbol{u}$). Essentially, the result "$^{n}\boldsymbol{y_d}$" is asserted to be *less* accurate than "$\boldsymbol{y_d}$". Thus, e.g. the error "$|{}^{n}Đ_1|$" is **≤2.17**% (cf. Table **3**); but (cf. Table **1**) "$|Đ_1|$" is **≤1.3**%. Further, as clarified by Table 4, uncertainty "$^{n}\boldsymbol{\varepsilon_1}$" is "2.26$^{G}\boldsymbol{u}$ = **2.26**%", i.e. though (cf. Table **2**)**:** $\boldsymbol{\varepsilon_1}$ **=** 1.40$^{G}\boldsymbol{u}$ = **1.40**%. In other words, the estimate as**:** $\boldsymbol{y_1}$ = –11.484×10$^{-3}$ (cf. example no **0** or **00** in Table **1**); rather than as $^{n}\boldsymbol{y}_1$ = –14.659×10$^{-3}$ (or**:** $^{n}\boldsymbol{y}_1$ = –14.638×10$^{-3}$, cf. Table **3**) should represent the desired differential $^{13/12}$**C** abundance ratio (***Y***$_1$) in the sample[9] ***S***. It is really thus signified (cf. Fig. 1***a*** and Fig. 1***b***) that different *source*/ *lab* specific results by the standardization method [viz.**:** $y_d^{\mathbf{Exp.1}}$, $y_d^{\mathbf{Exp.2}}$ **…** (***d*** = 1, or 2, or 3) in Table 1] should be more closely comparable than the corresponding those [$^{n}y_d^{\mathbf{Exp.1}}$, $^{n}y_d^{\mathbf{Exp.2}}$ **…** in Table 3] by the present method.

4.1.1.1 Can the process as Eq. 14 cause its output to be free from "***W***" (and accurate)?

Table 4 (cf. for Eq. 14) clarifies that the rate of variation, e.g. of $^{nI}Y_J$ with $X_J$, is decided by $X_J$ itself, i.e.**:** $^{nX}M_J^J = {}^{nX}f_J^J(X_J, Z1_J, Z2_J)$. Thus, as discussed in the standardization case above, the scale-converted estimate $^{nI}y_J$ **(**and hence any desired result $^{n}y_d$, and/ or *achievable* accuracies**:** $^{\mathrm{nI}}\varepsilon_J$ **…** $^{\mathrm{n}}\varepsilon_d$) should depend on, among others, the reference ***W*** used for measurements.

Further (Table 4)**:** $^{\mathrm{nI}}\varepsilon_J$ = 2.056$^{G}u$ [whereas**:** $^{\mathrm{I}}\varepsilon_J$ = 1.2$^{G}u$ (cf. above for Eq. **5a: (**$x_J$**,** $z1_J$**)** → $^{I}y_J$], i.e. $^{nI}y_J$ has to be less accurate than $^{I}y_J$. In other words, Eq. 14 itself is responsible for the inaccuracy of a desired result as $^{n}y_d$ [than as $y_d$]. In any case, Eq. 14 cannot cause a higher stage process to vary in nature from the standardization case, viz.**: (i)** like**:** $^{I}y_d \rightarrow {}^{II}y_d$, the "(*differential*) $^{nI}y_d$ → (*absolute*) $^{nII}y_d$" conversion causes **reduction** of uncertainty "$^{\mathrm{nII}}\varepsilon_d \ll {}^{\mathrm{nI}}\varepsilon_d$" [cf. Table 4 for Eq. 15]; **(ii)** the evaluated carbon and oxygen abundance ratios "$\{^{nIII}y_d\}_{d=1}^{3}$" appear, like the above case, equally accurate as their input isotopic $CO_2$ abundance ratios $^{nII}y_J$ and $^{nII}y_K$ [e. g.**:** $|{}^{\mathrm{nIII}}Đ_1| \approx |{}^{\mathrm{nII}}Đ_J|$, cf. Table 3; or**:** $^{\mathrm{nIII}}\varepsilon_1 \approx {}^{\mathrm{nII}}\varepsilon_J$, cf. Table 4]; and, **(iii)** the shaping of a result as a **δ**-value, $^{nIII}y_d \rightarrow {}^{n}y_d$, is again (i.e. like**:** $^{III}y_d \rightarrow y_d$) demonstrated to be an *error*-***enhancing*** task **(**viz. **[** $|{}^{\mathrm{n}}Đ_1| / |{}^{\mathrm{nIII}}Đ_1|$ **] ≈ 67**, cf. Table 3; and/ or**:** [$^{\mathrm{n}}\varepsilon_1/{}^{\mathrm{nIII}}\varepsilon_1$] = 66.5, cf. Table 4**)**.

4.1.1.2 Specific observations**:** isn't the output accuracy "ε" predictable?

It may be noted that (cf. Table **4** for Eq. 14)**:** $^{nX}M_i^d = -[{}^{nZ1}M_i^d + {}^{nZ2}M_i^d]$; and [6,19]**:** $^{\mathrm{nI}}Đ_d$ = [($^{nX}M_i^d \times {}^{X}\Delta_i$) + ($^{nZ1}M_i^d \times {}^{Z1}\Delta_i$) + ($^{nZ2}M_i^d \times {}^{Z2}\Delta_i$)], with**:** ***d* = *i* = *J***, or ***K***. That is, if**:** $^{X}\Delta_i = {}^{Z1}\Delta_i$ =

${}^{Z2}\Delta_i$, then: $^{nI}Đ_d = 0$; which explain why, only in the case of example no. 5 in Table 3, the errors $^{nI}Đ_d$ and $^{nII}Đ_d$ [i.e. ($^{nII}M_d^d \times {}^{nI}Đ_d$)] have turned out zero. Further, for equal but opposite input-errors (cf. example nos.1 and 2, or nos. 3 and 4), the output-errors are generally asymmetric (e.g. "$+^{nI}Đ_J$" = **2.077**, but "$-^{nI}Đ_J$"= **2.036**). These exemplify (cf. Eqs. 12-13b$'$) how really an *error* as $^{nI}Đ_d$, i.e. *uncertainty* $^{nI}\varepsilon_d$, and/ or *output* $^{nI}y_d$, should stand for ***systematic*** (here, Eq. **14** dictated) changes in respective *inputs*. In other words, these provide an insight into the fact that output accuracy $\varepsilon_d$ [to be achievable by any specified method of evaluation] is a priori predictable.[6]

4.1.1.3 Does Eq. 14 represent, at all, a scale conversion method?

The scale conversion variable is "$^{I}Y_d = ([R_i/{}^{D}R_i] - 1)$", cf. Eq. 5. Thus, if Eq. 14 should also represent a valid process as Eq. 5, then $^{nI}Y_d$ (and hence "$^{nII}Y_d$, $^{nIII}Y_d$ and $^{n}Y_d$") will be identical with $^{I}Y_d$ (and "$^{II}Y_d$, $^{III}Y_d$ and $Y_d$", respectively). Unfortunately, Eq. 14 cannot be derived from the fundamental principle indicated by Eq. 5/ 5a above. However, Eq. 14 can easily be translated to:

$$\frac{R_i}{{}^{D}R_i} = \left(\left[\frac{R_i}{{}^{W}R_i} - \frac{{}^{A1}R_i}{{}^{W}R_i}\right] \times \left[\frac{{}^{A2}R_i}{{}^{D}R_i} - \frac{{}^{A1}R_i}{{}^{D}R_i}\right] \Big/ \left[\frac{{}^{A2}R_i}{{}^{W}R_i} - \frac{{}^{A1}R_i}{{}^{W}R_i}\right]\right) + \frac{{}^{A1}R_i}{{}^{D}R_i} \tag{18}$$

Again, the right hand side (RHS) of Eq. 18 itself can be rewritten as:

$$\text{RHS} = \left(\frac{R_i}{{}^{W}R_i}\left[\frac{{}^{A2}R_i}{{}^{D}R_i} - \frac{{}^{A1}R_i}{{}^{D}R_i}\right] - \left[\frac{{}^{A1}R_i}{{}^{W}R_i} \times \frac{{}^{A2}R_i}{{}^{D}R_i}\right] + \left[\frac{{}^{A1}R_i}{{}^{D}R_i} \times \frac{{}^{A2}R_i}{{}^{W}R_i}\right]\right) \Big/ \left(\frac{{}^{A2}R_i}{{}^{W}R_i} - \frac{{}^{A1}R_i}{{}^{W}R_i}\right)$$

$$= \left(\frac{R_i}{{}^{W}R_i}\left[\frac{{}^{A2}R_i - {}^{A1}R_i}{{}^{D}R_i}\right] - \left[\frac{{}^{A1}R_i}{{}^{W}R_i} \times \frac{{}^{A2}R_i}{{}^{D}R_i}\right] + \left[\frac{{}^{A1}R_i}{{}^{D}R_i} \times \frac{{}^{A2}R_i}{{}^{W}R_i}\right]\right) \times \left(\frac{{}^{W}R_i}{{}^{A2}R_i - {}^{A1}R_i}\right)$$

Thus the requirement for the RHS to be reduced to the left hand side (as "$R_i/{}^{D}R_i$") is that:

$$\left[\frac{{}^{A1}R_i}{{}^{W}R_i} \times \frac{{}^{A2}R_i}{{}^{D}R_i}\right] = \left[\frac{{}^{A1}R_i}{{}^{D}R_i} \times \frac{{}^{A2}R_i}{{}^{W}R_i}\right] \tag{19}$$

Clearly, in terms of all individual $R_i$-values, Eq. 19 is valid. However: $([{}^{A1}R_i/{}^{W}R_i] \times [{}^{A2}R_i/{}^{D}R_i])$ $\neq$ $([{}^{A1}R_i/{}^{D}R_i] \times [{}^{A2}R_i/{}^{W}R_i])$. This is because that the ratios-of-ratios as "$^{A1}R_i/{}^{D}R_i$" and "$^{A2}R_i/{}^{D}R_i$" (and/ or the relative differences: $C1_i$ and $C2_i$, cf. Eq. 14) are *constants*, whereas $Z1_i$ and $Z2_i$ (and thus the ratios: "${}^{A1}R_i/{}^{W}R_i$" and "${}^{A2}R_i/{}^{W}R_i$") are *variables*. Therefore the **true** scale converted data, which is represented by "${}^{I}Y_d$" (cf. Eq. **5/ 5a**), cannot be identical with "${}^{nI}Y_d$". In other words, the use of Eq. **14** should cause the desired results to really be fictitious.

## 4.2 The expected normalization ($X_i \xrightarrow{A1,A2} {}^{I}Y_d$) method

As clarified by the scale conversion principle (cf. **Eq. 5**), two or even more than two different ***Ai***-standards could really be involved as follows:

$$\frac{R_i}{{}^{D}R_i} = \frac{R_i}{{}^{W}R_i} \times \frac{{}^{W}R_i}{{}^{A2}R_i - {}^{A1}R_i} \times \frac{{}^{A2}R_i - {}^{A1}R_i}{{}^{D}R_i} = \left(\frac{R_i}{{}^{W}R_i} \times \left[\frac{{}^{A2}R_i}{{}^{D}R_i} - \frac{{}^{A1}R_i}{{}^{D}R_i}\right]\right) \Big/ \left(\frac{{}^{A2}R_i}{{}^{W}R_i} - \frac{{}^{A1}R_i}{{}^{W}R_i}\right) \qquad (20)$$

Or,

$$\left(\frac{R_i}{{}^{D}R_i} - 1\right) = \frac{\left(\left[\frac{R_i}{{}^{W}R_i} - 1\right] + 1\right) \times \left(\left\{\left[\frac{{}^{A2}R_i}{{}^{D}R_i} - 1\right] + 1\right\} - \left\{\left[\frac{{}^{A1}R_i}{{}^{D}R_i} - 1\right] + 1\right\}\right)}{\left(\left[\frac{{}^{A2}R_i}{{}^{W}R_i} - 1\right] + 1\right) - \left(\left[\frac{{}^{A1}R_i}{{}^{W}R_i} - 1\right] + 1\right)} - 1$$

That is,

$${}^{I}Y_d = \left((X_i + 1) \times \frac{(C2_i - C1_i)}{(Z2_i - Z1_i)} - 1\right), \quad d = i = J, K \qquad (5b)$$

Thus, Eq. 5b must be a valid scale conversion formula. Further, in terms of standards (true values of the variables: $X_i$, $Z1_i$, and $Z2_i$), it could be shown that the outputs of Eq. **5**, Eq. **5a** and Eq. **5b** are identical. Therefore, the 2nd, 3rd and 4th stage processes: ${}^{I}Y_d \rightarrow {}^{II}Y_d \rightarrow {}^{III}Y_d \rightarrow Y_d$ cannot be different from those as Eq. **6**, Eq. **7** and Eq. **8**, respectively.

### 4.2.1 *Can a "$X_i \xrightarrow{A1,A2} {}^{I}Y_d$" process, even Eq. 5b, help yield accurate results?*

We consider $X_i$, $Z1_i$, and $Z2_i$ in Eq. 5b to be represented by their measured[9] *estimates* used (for Eq. 14/ 6a) above. The results thus obtained are described as example no. 00 in Table 5 **(**but for distinguishing the present COCP [i.e. Eqs. **5b**, **6**, **7** and **8**] specific outputs from those by the COCPs discussed above**)** using the *notation-prefix* "***e***". However any estimate in Table 5 is, it may be pointed out, quite different from the corresponding estimate in either Table 1 or 3.

4.2.1.1 Is the method as Eq. 5b better than that as Eq. 5a/ 14?

Example nos. 1-5 in Table 5, and Fig. 1c, clarify how the results by the present method may vary between labs of a *given* measurement-accuracy ($^{G}\boldsymbol{u}$), viz. ±**1**%, and ±**0.5**%, respectively. However, the outputs (Table 5) by Eq. 5b itself are largely varied, viz. (errors)**:** $^{\mathbf{Max}}|^{\mathbf{eI}}\mathbf{Đ_J}| \approx 4.6^{G}\boldsymbol{u}$, and**:** $^{\mathbf{Max}}|^{\mathbf{eI}}\mathbf{Đ_K}| \approx 273^{G}\boldsymbol{u}$. These errors are, of course (like the above cases), reduced by the 2nd stage (Eq. 6)**:** $^{\mathbf{Max}}|^{\mathbf{eII}}\mathbf{Đ_J}| = 1.28^{G}\boldsymbol{u}$, and**:** $^{\mathbf{Max}}|^{\mathbf{eII}}\mathbf{Đ_K}| = 2.25^{G}\boldsymbol{u}$. Yet, the latter errors are much **higher** than the respective standardization errors (i.e. for using [Eqs. 5a and 6, or alone] Eq. 6a, cf. Table 1)**:** $^{\mathbf{Max}}|^{\mathbf{II}}\mathbf{Đ_J}| = 0.014^{G}\boldsymbol{u}$, and**:** $^{\mathbf{Max}}|^{\mathbf{II}}\mathbf{Đ_K}| = 0.021^{G}\boldsymbol{u}$. Further, the behavior-governing parameters (cf. Eqs. 9-11) of the present COCP are described in Table 6, which confirm Eq. 5b to represent a rather *undesirable* scale conversion method, e.g.**:** $^{\mathbf{eI}}\boldsymbol{\varepsilon}_{\mathbf{J}} = \mathbf{4.5}^{\mathbf{G}}\mathbf{u}$ (i.e. though**:** $^{\mathbf{I}}\boldsymbol{\varepsilon}_{\mathbf{J}} = \mathbf{1.2}^{\mathbf{G}}\mathbf{u}$ [cf. above for Eq. 5a]; and**:** $^{\mathrm{nI}}\boldsymbol{\varepsilon}_{\mathbf{J}} = \mathbf{2.056}^{\mathbf{G}}\mathbf{u}$ [cf. Table 4 for Eq. 14]).

The scale-conversion by employing two ***Ai***-standards will in any case (cf. Eq. 5b or Eq. 14) require *two **additional*** (viz. $\boldsymbol{Z1_J}$ and $\boldsymbol{Z2_J}$) measurements, and that by using single ***Ai*** (cf. Eq. 5a) will involve only *one* additional ($\boldsymbol{Z1_J}$) measurement. Thus, in general, any estimate as $^{eI}\boldsymbol{y_J}$ (cf. Eq. **5b**) or $^{nI}\boldsymbol{y_J}$ (cf. Eq. 14) should be subject to larger error than $^{I}\boldsymbol{y_J}$ (cf. Eq. **5a**).

4.2.1.2 Can Eq. 5b make its output free of the measurement-reference "***W***"?

The rate of variation of e.g. "$^{eI}\boldsymbol{Y_J}$ with $\boldsymbol{X_J}$" is governed as (cf. Table 6 for Eq. **5b**)**:** $^{eX}\boldsymbol{M_J^J} = {}^{eX}\boldsymbol{f_J^J}(\boldsymbol{X_J}, \boldsymbol{Z1_J}, \boldsymbol{Z2_J})$. Therefore, the estimate as $^{eI}\boldsymbol{y_J}$ (and thus the desired result $^{e}\boldsymbol{y_d}$), i.e. achievable accuracy $^{\mathbf{eI}}\boldsymbol{\varepsilon}_{\mathbf{J}}$ (and in turn, $^{\mathbf{e}}\boldsymbol{\varepsilon}_{\mathbf{d}}$), should be decided by the measured variables ($\boldsymbol{X_J}$, $\boldsymbol{Z1_J}$, and $\boldsymbol{Z2_J}$), and hence by the working reference ***W***.

4.2.1.3 Can scale conversion cause a **δ**-result to be more accurate than its absolute value?

It should be of interest to note that the different evaluation-methods are indistinguishable, in nature, from one another [comparison between Fig. 1**a**, Fig. 1**b** and Fig. 1**c**]. For example, the translation of a **δ**-estimate as even "$^{eI}y_d$" into its absolute value "$^{eII}y_d$" is (like: $^{I}y_d$ into $^{II}y_d$; or: $^{nI}y_d$ into $^{nII}y_d$) shown to improve accuracy, viz.: $^{eII}\varepsilon_J$ = ([1.26$^{G}$**u**]/[4.5$^{G}$**u**])$^{eI}\varepsilon_J$ = 0.28$^{eI}\varepsilon_J$ (cf. Table 6). This explains (cf. Fig. **1c**) why should a **4.6**% variation (between the scale converted data $^{eI}y_J^{Lab1}$ and $^{eI}y_J^{Lab2}$) be reduced to **1.3**% (between the evaluated isotopic-$CO_2$ abundance ratios $^{eII}y_J^{Lab1}$ and $^{eII}y_J^{Lab2}$). Moreover, the conversion of e.g. the *absolute* ($^{18}O/^{16}O$ abundance) ratio, $^{eIII}y_3$, into the desired *differential* estimate, $^{e}y_3$, is (like: $^{III}y_3$ into $y_3$; or: $^{nIII}y_3$ into $^{n}y_3$) demonstrated to cause the ***compromising*** of achievable accuracy and/ or comparability (cf. Table 6: ($^{e}\varepsilon_3/^{eIII}\varepsilon_3$) = **109.8** or, Table 5: ( | $^{e}Đ_3$ | / | $^{eIII}Đ_3$ | ) ≈ **110)**. This clarifies why a relatively true variation of **2.3**% between *two* source/ lab specific results (cf.: $^{eIII}y_3^{Lab1}$ and $^{eIII}y_3^{Lab2}$ in **Fig. 1c**) might be mistaken to be as ≈**248**% (cf. $^{e}y_3^{Lab1}$ and $^{e}y_3^{Lab2}$).

Let us now say that three different labs had used the three different data normalization methods discussed here, and reported the results as the example **no. 00** in Tables **1**, **3** and **5**. Then, it may be seen that the scatter between the desired *differential* lab-results ($y_3$, $^{n}y_3$ and $^{e}y_3$) is as high as **41.3**%, whereas that between their *absolute* values ($^{III}y_3$, $^{nIII}y_3$ and $^{eIII}y_3$) is **0.72**%. That is, absolute estimates as even: $^{III}y_3$, $^{nIII}y_3$ and $^{eIII}y_3$ are far more ***closely*** intercomparable than their desired "**δ**" estimates as: $y_3$, $^{n}y_3$ and $^{e}y_3$, however.

## 5. ARE THE ABOVE FINDINGS A SOURCE-SPECIFIC?

The variations of stage specific *outputs* by all three above discussed COCPs as a function of the *measured variable* "$X_J$" (i.e. with sample ***source***) are depicted in Fig. 2. However, a change of source should mean variations in both "$X_J$" and "$X_K$". In addition, the *auxiliary*-variables $Zi_J$ and $Zi_K$ may themselves be varied, i.e. for lab specific choice of $Ai$-standard(s) and/ or working reference $W$. Above all, any measurement is subject to error. Thus, while the variation of even alone a source should get signified by *a net change* in any stage/ COCP output, the latter may turn out to represent ***multiple*** causes of input-variations. This is the reason why, even for a given source $S$ (as ref. [9]) and the measurement-accuracy $^{G}u$ (as 1%), the different possible method specific output-variations are in Tables 1, 3 and 5 numerically exemplified. Similarly, for Fig **2**, (the true values of) all method-governing **other** *variables* [i.e. barring the "$X_J$"], and (thus even the working-reference $W$, and) all *constants*, are considered to be the same as those[9] mentioned above. However, in terms of **carbon** alone, any variation in source could be shown[21] to imply the variations of essentially (the isotopic $^{44}[CO_2]$ and $^{45}[CO_2]$ abundances, i.e. of) the abundance-ratio $R_J$, and hence "$X_J$", only. In other words, a given $X_J$ should really represent a given carbon source as that indicated by the top X-**axis** in Fig. 2(**I**), which gives the $^{13}$**C**-abundances as those obtained by the standardization-COCP.

However, a variation in $X_J$ should, in turn, introduce**:** **(i)** a net change **in** the scale converted output $^{I}Y_J$; **(ii)** a change [***due to*** $^{I}Y_J$] in the 2nd stage output $^{II}Y_J$; and **(iii)** hence, changes in all 3rd and 4th stage outputs ($^{III}Y_d$ and $Y_d$, respectively, with**:** $d$ = 1, 2 and 3). Thus, although [for a given method] the different *stage* outputs **(**viz.**:** $^{eI}Y_J$, $^{eII}Y_J$, $^{eIII}Y_1$, and $^{e}Y_1$**)** are ever different from one another, the 2nd-4th stage curves (cf. Fig. 2**(II-IV)**) must be parallel to the corresponding 1st stage curve (cf. Fig. 2**(I)**). However, all three method specific curves are *displaced* [e.g. in Fig. 2(I)] from one another, which confirm that any stage/ COCP accuracy "**ε**" (cf. Tables 2, 4 and 6) is *method-dependent*. For illustration, let's <u>assume</u> that the conventional normalization (**C. N.**) method is valid. Then it may be noted that an estimate by the "**C. N.**" method, can though turn out to be equally accurate as, cannot be more accurate than, the corresponding estimate by the standardization (**STN.**) method [comparison of *slopes* between e.g. "$^{nI}Y_J$- and $^{I}Y_J$" curves].

Further, the **uncertainty-factors** for evaluating the different *source* and *stage* specific results [as Fig. 2] are described in Fig. **3**, which supplements the above findings that**: (i)** any stage or COCP estimate by the **STN.-**method should *best* represent the corresponding source, and**: (ii)** the largely different outputs **(**i.e. those by the expected normalization [**E. N.**] method) should be *least* accurate. For illustration, we consider a specific *source* as**:** $\boldsymbol{X_J}$ = –0.02, and assume the achievable accuracy ($^{G}\boldsymbol{u}$) for *any* measurement to be 0.01%. Then, as predicted by Fig. 3(**IV**), the desired result as "$^{n}\boldsymbol{y_1}$" should be **(**$^{n}\boldsymbol{\varepsilon_{13/12}}$ **= (**$^{n}\boldsymbol{[UF]_{13/12}}$ × $^{G}\boldsymbol{u}$**) = (2.13 × 0.01) ≈)** 0.02% accurate. However, the result as "$\boldsymbol{y_1}$" [cf. the **STN.-**curve in Fig **2(IV)**] should turn out (cf. Fig. **3(IV)**) *more* accurate**:** $\boldsymbol{\varepsilon_{13/12}}$ = ($\boldsymbol{[UF]_{13/12}}$ × $^{G}\boldsymbol{u}$) = (**1.21 × 0.01**) ≈ 0.01%. Moreover, the result (by the **E. N.** method) $^{e}\boldsymbol{y_1}$ should really, i.e. like the above exemplified source [cf. Tables 1- 6], be ***least*** accurate**:** $^{e}\boldsymbol{\varepsilon_{13/12}}$ = ($^{e}\boldsymbol{[UF]_{13/12}}$ × $^{G}\boldsymbol{u}$) = (**4.85 × 0.01**) ≈ 0.05%.

### 5.1 Shouldn't absolute [rather than δ] difference be the source-characterizing key?

As clarified in Fig. 3, any evaluated isotopic-$CO_2$ abundance ratio as $^{II}\boldsymbol{y_J}$ [or a constituent-elemental isotopic-ratio as $^{III}\boldsymbol{y_1}$] should, even irrespective of scale conversion method, be ***more*** accurate than the corresponding *differential* estimate $^{I}\boldsymbol{y_J}$ [or $\boldsymbol{y_1}$]. Thus if (e.g.) "$\boldsymbol{X_J}$ = –0.02" is considered to be an unknown source "***S*1**", and "**STN.**" is the method of evaluation (i.e. if e.g. [true]**:** "$^{IIIS1}\boldsymbol{Y_1} \equiv {}^{S1}\boldsymbol{E_{13/12}}$" = 10.994891×10$^{-3}$; and/ or**:** "$^{S1}\boldsymbol{Y_1} \equiv {}^{S1/D}\boldsymbol{\delta_{13/12}}$" = –21.56309×10$^{-3}$); then a determined absolute result $^{IIIS1}\boldsymbol{y_1}$ is predicted to be "($\boldsymbol{[UF]_{13/12}}$ / $^{III}\boldsymbol{[UF]_{13/12}}$) = (1.21 / 0.027) ≈ **45**" [cf. Fig. 3(**III-IV**)] times more accurate than the desired "$^{S1/D}\boldsymbol{\delta_{13/12}}$" value ($^{S1}\boldsymbol{y_1}$).

However, say that actual measurement of **S1** has yielded**:** $\boldsymbol{x_J}$ = ($\boldsymbol{X_J}$ + [–0.01%]) = –0.019998, thereby giving **(i):** $^{IIIS1}\boldsymbol{y_1}$ = 10.994915×10$^{-3}$ = ($^{S1}\boldsymbol{E_{13/12}}$ + $^{III}\boldsymbol{Đ_1}$) = (10.994891×10$^{-3}$ + 0.00022%); and **(ii):** $^{S1}\boldsymbol{y_1}$ = –21.56095×10$^{-3}$ = ($^{S1/D}\boldsymbol{\delta_{13/12}}$ + $\boldsymbol{Đ_1}$) = (–21.56309×10$^{-3}$ – 0.01%). Thus, either the absolute estimate $^{IIIS1}\boldsymbol{y_1}$; or even the "[$\boldsymbol{Đ_1}$ / $^{III}\boldsymbol{Đ_1}$] = [0.01 / 0.00022] ≈ **45**" times more erroneous

**δ**-estimate (i.e. $^{S1}y_1$), reflects ***S*1** to be (with reference to the reference standard ***D***) a ***depleted*** source. Further, the *absolute* difference ($\Delta^{S1}E_{13/12}$ = [$^{IIIS1}y_1 - {}^{D}E_{13/12}$] = [$^{IIIS1}y_1 - 11.2372\times10^{-3}$] = **$-0.2423\times10^{-3}$**) indicates ***S*1** to be somewhat close to the standard ***D*** [in terms of isotopic composition (IC)]. However, "$^{S1}y_1$ = **$-21.56095\times10^{-3}$** ≈ (89 × $\Delta^{S1}E_{13/12}$)", i.e. the **δ**-difference should mislead one to consider ***S*1** to be, by its IC, very different from the source as ***D***.

**5.2 Can any δ-result turn out to be absurd but the absolute result as highly accurate?**

Each curve in Fig. 3 has a peak or a valley **(**cf., for a typical **E. N.** case, the insert in Fig. 3(**I**)**)**. However why at all $^{I}[UF]_J$ (i.e. scale [$X_J \rightarrow {}^{I}Y_J$, i.e.**:** $^{S/W}\delta_{45/44} \rightarrow {}^{S/D}\delta_{45/44}$] conversion *accuracy* $^{I}\varepsilon_J$) should vary as a function of $X_J$ (*sample*) is that "$^{I}[UF]_J$" (or $^{nI}[UF]_J$, or so) is, one may verify, controlled by the inverse-factor "**(**$^{S}R_J - {}^{D}R_J$**)**$^{-1}$". Therefore, if**:** "($^{S}R_J - {}^{D}R_J$) → **0**", then "$^{I}[UF]_J$" will tend to be *infinity*, i.e. any curve as Fig. 3(**I**) should pass through a maxima.

Further, as shown in Table 2/ 4/ 6, the *rate* ($M_1^1$) of variation of $Y_1$ as a function of $^{III}Y_1$ (i.e. of $^{S/D}\delta_{13/12}$ with $E_{13/12}$) is governed by the ratio**:** ($^{III}Y_1$ / [$^{III}Y_1 - {}^{D}E_1$]). Thus, for any conceivable case, $|M_1^1|$ is >**1**. Moreover, "**(**$^{III}Y_1 - {}^{D}E_1$**)** → **0**" should imply "$|M_1^1| \rightarrow \infty$". That is (cf. Eq. **13b**) the "$[UF]_{13/12}$ vs $X_J$" variation (cf. Fig. 3(**IV**)) will have [like Fig. 3(**I**)] a maxima.

However, what is thus signified is that **[**though a measured data as $x_J$ (i.e. $^{S/W}\delta_{45/44}$) may, for IC of the sample ***S*** to be *close* to IC of the working-reference ***W***, turn out to be *accurate***]** the scale-converted data as "$^{I}y_J$" (i.e. $^{S/D}\delta_{45/44}$) and, in turn, any desired result as "$y_1$" (i.e. $^{S/D}\delta_{13/12}$), will be increasingly ***inaccurate*** (cf. Fig. 3(I) and Fig. 3(IV), respectively) for IC of the sample ***S*** to be increasingly ***closer*** to IC of the reference standard ***D***.

Again, *decreasing* "$^{III}[UF]_1$" (cf. Fig. 3(III)) should generally cause "$[UF]_1$" (cf. Fig. 3(IV)) to be *increasing*. That is, a more **accurate** absolute estimate $^{III}y_1$ may yield a more **inaccurate** differential estimate $y_1$. However, why it is so? —— We recollect that (COCP-uncertainty, cf.

Eq. 12b)**:** $\boldsymbol{\varepsilon_1 = ([UF]_1 \times {}^{G}u)}$ = ([$|\boldsymbol{M_1^1}|$ × $^{III}[UF]_1$] × $^{G}\boldsymbol{u}$). Further, by**:** $\boldsymbol{({}^{III}Y_1 - {}^{D}E_1) \to 0}$, it is meant that**: (i)** "$^{III}[UF]_1$" to be reducing [cf. Fig. 3(**III**)], but**: (ii)** "$|\boldsymbol{M_1^1}| \to \infty$" (cf. above). That is, "$|\boldsymbol{M_1^1}|$" overrides "$^{III}[UF]_1$" in governing "$[UF]_1$". However, the interesting point is that a *source* as "$\boldsymbol{({}^{S}E_1 - {}^{D}E_1) \approx 0}$" should (though, in terms of absolute estimate, be **accurately** represented; cf. Fig. 3(**III**)) really be ***misrepresented*** by corresponding **δ-**result [cf. Fig. 3(**IV**)].

However, why Fig. 3(**II**) should at all have a minima is that the evaluation of any absolute value [here**:** $\boldsymbol{{}^{II}Y_J = {}^{II}f_J({}^{I}Y_J)}$, i.e.**:** $\boldsymbol{{}^{S}R_{45/44} = {}^{II}f_J({}^{S/D}\delta_{45/44})}$] is a desirable task (cf. Table 2/ 4/ 6 for 1st C. stage)**:** $\boldsymbol{|{}^{II}M_J^J| < 1}$, and/ or**:** $\boldsymbol{{}^{II}[UF]_J < {}^{I}[UF]_J}$ (i.e. as**:** $\boldsymbol{{}^{II}[UF]_J}$ = ([ $|{}^{II}\boldsymbol{M_J^J}|$ × $^{I}[UF]_1$]). Further, "$^{II}\boldsymbol{M_J^J}$" could be shown to equal the ratio as "([$^{S}\boldsymbol{R_J} - {}^{D}\boldsymbol{R_J}$] / $^{S}\boldsymbol{R_J}$)". Thus, "$\boldsymbol{({}^{S}R_J - {}^{D}R_J) \to 0}$" should not only mean**:** $\boldsymbol{{}^{I}[UF]_J \to \infty}$ (cf. above) but also**:** $\boldsymbol{|{}^{II}M_J^J| \to 0}$. Clearly, here, "$^{II}\boldsymbol{M_J^J}$" dominates over "$^{I}[UF]_J$" in deciding "$^{II}[UF]_J$" and/ or the uncertainty "$^{II}\boldsymbol{\varepsilon_J}$".

Moreover, Table 2/ 4/ 6 also clarifies e.g. that**:** $\boldsymbol{|{}^{III}M_\beta^1| \ll 1}$, which in turn explains [cf. Fig. 3(II) and Fig. 3(III)] why**:** $\boldsymbol{{}^{III}[UF]_1 \approx {}^{II}[UF]_J}$.

## 5.3 Isn't it possible to a priori authenticate a method of evaluation?

It is clarified above that uncertainty-factor ***[UF]***, or uncertainty **ε**, for evaluating any output "***Y***", is (like ***Y*** itself) *systematic* by nature, i.e. governed by input-output relation(s) yielding ***Y***.[6] Further, that this is a fact is supplemented by the observation that the curves as Fig. 2 and Fig. 3 are ***asymmetric*** around "$\boldsymbol{X_J = 0}$". In other words, Fig. **3** exemplifies how it should in practice be possible to a priori ascertain [viz. as**:** $\boldsymbol{\varepsilon = ([UF] \times {}^{G}u)}$] whether a given *evaluation-model* (here, the scale-conversion by employing *one* or *two* different ***A*i-**standards; and/ or the determining of a result as **δ-** than as *absolute*-value) meets the purpose it is designed for.

## 6. CONCLUSIONS

The above study clarifies that any *directly* (or even *indirectly*) measured **δ**-estimate should be rather ***inaccurate*** than its evaluated absolute value. For example, "$^{S}\boldsymbol{R}_{i}^{CO_2} \equiv {}^{II}\boldsymbol{Y}_i$" than "$^{S/W}\boldsymbol{\delta}_{i}^{CO_2} \equiv \boldsymbol{X}_i$" (and "$^{S}\boldsymbol{E}_d \equiv {}^{III}\boldsymbol{Y}_d$" than "$^{S/D}\boldsymbol{\delta}_d \equiv \boldsymbol{Y}_d$") values are shown to turn out, irrespective of **δ**-scale conversion method, less erroneous. That is, as signified above, different lab/ source specific absolute results (e.g.: $^{III}\boldsymbol{y}_d^{S1}$, $^{III}\boldsymbol{y}_d^{S2}$ **…**) should better represent their sources (***S*1**, ***S*2** **…**), and/ or be more closely intercomparable, than their **δ**-estimates ($\boldsymbol{y}_d^{S1}$, $\boldsymbol{y}_d^{S2}$**…**). In other words, *absolute* differences (as: [$^{III}\boldsymbol{y}_d^{S1} - {}^{D}\boldsymbol{E}_d$], [$^{III}\boldsymbol{y}_d^{S2} - {}^{D}\boldsymbol{E}_d$] **…**) rather than **δ**-differences ($\boldsymbol{y}_d^{S1}$, $\boldsymbol{y}_d^{S2}$**…**) should be the keys for identifying [and thus expounding the causes of] source-variations.

Further, achievable accuracy "**ε**" (for determining any stage/ COCP result "***Y***") is shown to be controlled by the ***difference**-in-isotopic-composition* between the sample ***S*** and the reference-standard ***D***. However the important finding is that, if ***S*** should be [increasingly] close to ***D***, then all evaluated *absolute* ($^{S}\boldsymbol{R}_{i}^{CO_2}$ and $^{S}\boldsymbol{E}_d$) values will really turn out [increasingly] *accurate*, but all <u>corresponding **δ**</u> (i.e. $^{S/D}\boldsymbol{\delta}_{i}^{CO_2} \equiv {}^{I}\boldsymbol{Y}_i$, and $^{S/D}\boldsymbol{\delta}_d \equiv \boldsymbol{Y}_d$) values should be [increasingly] ***inaccurate***.

Thus, though: **ε = (*[UF]* × *u*)**,[6,23] the **ε**-value is shown above to be decided by the uncertainty-factor "***[UF]***" (i.e. *nature* of input-output relation(s) yielding "***Y***") rather than by the achievable measurement-accuracy "***u***". It is clarified how the knowledge of *any stage* (or *COCP*: "$^{S/W}\boldsymbol{\delta}_{i}^{CO_2} \rightarrow {}^{S/D}\boldsymbol{\delta}_{i}^{CO_2} \rightarrow {}^{S}\boldsymbol{R}_{i}^{CO_2} \xrightarrow{\beta} {}^{S}\boldsymbol{E}_d \rightarrow {}^{S/D}\boldsymbol{\delta}_d$" ≡ "$\boldsymbol{X}_i \rightarrow {}^{I}\boldsymbol{Y}_i \rightarrow {}^{II}\boldsymbol{Y}_i \xrightarrow{\beta} {}^{III}\boldsymbol{Y}_d \rightarrow \boldsymbol{Y}_d$") *specific* ***[UF]*** can be acquired and checked, even a priori, whether the corresponding stage-process (or COCP) should meet the purpose it is designed for. For example, the outputs of the basic $CO_2^+$-IRMS evaluation "($^{II}\boldsymbol{Y}_J$, $^{II}\boldsymbol{Y}_K$, **β**) → {$^{III}\boldsymbol{Y}_d$}" were, in fact,[22] previously pointed out to be much less sensitive towards **β** than towards $^{II}\boldsymbol{Y}_J$ or $^{II}\boldsymbol{Y}_K$. However, this finding is supplemented above by demonstrating that, even for **±1**% variation in (*the* 3rd *stage input*) "**β**", the estimates "{$^{III}\boldsymbol{y}_d$}" remain as accurate as "$^{II}y_J$ and $^{II}y_K$". However, any process of the type "$^{S}E_d \rightarrow {}^{S/D}\delta_d$" ≡ "$^{III}Y_d \rightarrow Y_d$" is shown to imply the corresponding "$[UF]_d$" as >**1**, which explains why the **δ**-estimate "$y_d$" should be *less* accurate than the absolute-value "$^{III}y_d$". Similarly, the **reverse** kind of process (viz.**:** "$^{S/D}\delta_i^{CO_2} \rightarrow {}^{S}R_i^{CO_2}$" ≡ "$^{I}Y_i \rightarrow {}^{II}Y_i$") is shown to be characterized by**:** $[UF]_i < 1$. This explains why "($^{I}y_i \pm {}^{I}\varepsilon_i$) → ($^{II}y_i \pm {}^{II}\varepsilon_i$)" is observed above to imply**:** $^{II}\varepsilon_i < {}^{I}\varepsilon_i$.

However, depending upon the number of auxiliary reference standards ***Ai*** ($i$ = **1**, **2 …**) employed in the 1st stage process ("$^{S/W}\delta_i^{CO_2} \xrightarrow{Ai(s)} {}^{S/D}\delta_i^{CO_2}$" ≡ "$X_i \xrightarrow{Ai(s)} {}^{I}Y_i$"), the "$[UF]_i$" is shown above to vary. Further, the aid of even only ***A*1** means the involving of an *additional* ***measured*** data [($^{A1/W}\delta_i^{CO_2} \pm {}^{A1}u_i$) ≡ ($z1_i \pm {}^{Z1}u_i$)] in the process**:** $x_i \xrightarrow{z1_i} {}^{I}y_i$. That is the employing of ***Ai(s)*** cannot be a general means for ensuring "$^{I}y_i$" to be accurate. In any case, the idea that the scale-conversion with the aid of two ***Ai***-standards would cause $^{I}y_i$ to be more accurate than that for employing any single ***Ai***,[3,9,15,17] is verified above to stand for no general fact.

It is also pointed out above that the use of ***Ai(s)***, for scale conversion, cannot even cause $^{I}y_i$ (i.e. estimated $^{S/D}\delta_i^{CO_2}$, and hence the desired result $^{S/D}\delta_d$) to be independent from corresponding measurement-reference (***W***). That is, if should the measured data **(**for a sample ***S*** and given ***Ai***(s)**)** from two different labs be equally accurate but vary for using different references as "***W***" *only*, then also the *estimates* of $^{S/D}\delta_i^{CO_2}$ (and thus of $^{S}R_i^{CO_2}$, $^{S}E_d$ and $^{S/D}\delta_d$); and/ or *accuracy* $^{I}\varepsilon_i$ (and hence the *accuracies***:** $^{II}\varepsilon_i$, $^{III}\varepsilon_d$ and $\varepsilon_d$); should ***vary*** from one lab to another.

## ACKNOWLEDGEMENTS

The author is thankful to An Honorable Reviewer of RCM and Dr. A. K. Dhara for their valuable suggestions on the manuscript.

## REFERENCES

[1] J. T. Brenna, T. N. Corso, H. J. Tobias, R. J. Caimi. High precision continuous-flow isotope ratio mass spectrometry. *Mass Spectrom. Rev.* **1997**, *16*, 227.

[2] T. B. Coplen. Reporting of stable carbon, hydrogen, and oxygen isotopic abundances. *Pure and Applied Chem.* **1994**, *66*, 273.

[3] T. B. Coplen, W. A. Brand, M. Gehre, M. Groning, H. A. Meijer, B. Toman, R. M. Verkouteren. New guidelines for $\delta^{13}C$ measurements. *Anal. Chem.* **2006**, *78*, 2439.

[4] B. P. Datta, P. S. Khodade, A. R. Parab, A. H. Goyal, S. A. Chitambar, H. C. Jain. Molecular ion beam method of isotopic analysis: effect of error propagation, a case study with $Li_2BO_2^+$. *Rapid Commun. Mass Spectrom.* **1993**, *7*, 581.

[5] B. P. Datta, A. R. Parab, P. S. Khodade, S. A. Chitambar, H. C. Jain. Isotopic analysis of a $^6Li$ enriched lithium sample employing the $Li_2BO_2^+$ ion beam method: verification of the theoretical accuracy. *Int. J. of Mass Spectrom. and Ion Proc.* **1995**, *142*, 69.

[6] B. P. Datta. The theory of uncertainty for derived results: properties of equations representing physicochemical evaluation systems. *arXiv*: 0712:1732 [*physics.data-an*] **2007**.

[7] H. Craig. Isotopic standards for carbon and oxygen and correction factors for mass-spectrometric analysis of carbon dioxide. *Geochim. Cosmochim. Acta* **1957**, *12*, 133.

[8] C. E. Allison, R. J. Francey, H. A. Meijer. Recommendations for the reporting of stable isotope measurements of carbon and oxygen in $CO_2$ gas. *IAEA-TECDOC*-825, IAEA, Vienna, **1995**, pp. 155-162.

[9] R. M. Verkouteren, J. N. Lee. Web-based interactive data processing: application to stable isotope metrology. *Fresenius J. Anal. Chem.* **2001**, *370*, 803.

[10] W. G. Mook, P. M. Grootes. The measuring procedure and corrections for the high-precision analysis of isotopic abundance ratios, especially referring to carbon, oxygen and nitrogen. *Int. J. Mass Spectrom. Ion Phys.* **1973**, *12*, 273.

[11] K. J. Santrock, S. A. Studley, J. M. Hayes. Isotopic analyses based upon the mass spectrum of carbon dioxide. *Anal. Chem.* **1985**, *57*, 1444.

[12] S. S. Assonov, C. A. M. Brenninkmeijer. On the $^{17}O$ correction for $CO_2$ mass spectrometric isotopic analysis. *Rapid Commun. Mass Spectrom.* **2003**, *17*, 1007.

[13] R. M. Verkouteren, G. A. Klouda, L. A. Currie. The carbon dioxide isotopic measurements process: Progress at NIST on measurements, Reduction algorithms and standards, *IAEA-TECDOC*-825, IAEA, Vienna, **1995**, pp. 111-129.

[14] R. A. Werner, W. A. Brand. Review: Referencing strategies and techniques in stable isotope ratio analysis. *Rapid Commun. Mass Spectrom.*, **2001**, *15*, 501.

[15] D. Paul, G. Skrzypek, I. Forizs. Normalization of measured stable isotopic compositions to isotope reference scales — a review. *Rapid Commun. Mass Spectrom.* **2007**, *21*, 300.

[16] G. Skrzypek, R. Sadler, D. Paul. Error propagation in normalization of stable isotpe data: a Monte carlo analysis. *Rapid Commun. Mass Spectrom.* **2010**, *24*, 2697.

[17] R. Gonfiantini. Standards for stable isotope measurements in natural compounds. *Nature* **1978**, *271*, 534.

[18] ISO, Guide to the Expression of Uncertainty in Measurement. *ISO*, **1995**.

[19] J. B. Scarborough. *Numerical Mathematical Analysis*, Oxford & IBH Publishing Co., Kolkata **1966**.

[20] B. P. Datta. Theoretical evaluation of error in the isotopic analysis of carbon and oxygen as $CO_2^+$: considerations for determining two different isotopic abundance ratios simultaneously. *Rapid Commun. Mass Spectrom.* **2001**, *15*, 1346.

[21] B. P. Datta. Polynomial method of molecular isotopic abundance calculations: a computational note. *Rapid Commun. Mass Spectrom.* **1997**, *11*, 1767.

[22] B. P. Datta. Error in the solutions of a set of equations representing an experimental system: a case study for the simultaneous determination of $^{13}C/^{12}C$, $^{17}O/^{16}O$ and $^{18}O/^{16}O$ abundance ratios as $CO_2^+$. *Int. J. Mass Spectrom.* **2004**, *237*, 135.

[23] B. P. Datta. Uncertainty factors for stage-specific and cumulative results of indirect measurements. *arXiv*: 0909:1651 [*physics.data-an*] **2009**.

## APPENDIX A: Notations

**(**For clarity, *measured*-***sample*** and -***auxiliary*** variables are distinguished as "$X_i$" and "$Zi_i$ [viz. $Z1_i$, $Z2_i$ …]" respectively. Further, we refer to a **1st** stage output variable as $^{I}Y_d$ and, so.
In general**:** $i = J$ = "45/44", and**:** $i = K$ = "46/44".
Similarly**:** $d$ = 1 = "13/12", $d$ = 2 = "17/16" and $d$ = 3 = "18/16".
The abbreviation "COCP" generally refers to the *standardization* method [Eqs. **6a**, **7** and **8**].
The cases as the *conventional normalization* [COCP**:** Eqs. **14-17**] and the *expected normalization* {COCP**:** Eqs. **5a**, **6**, **7** and **8**] are in the text distinguished by additionally prefixing the symbols with "*n*" and "*e*" [i.e. as "$^{n}Y_3$", "$^{n}\varepsilon_3$" **…** and "$^{e}Y_3$", "$^{e}\varepsilon_3$" **…**], respectively.**)**

***Ai*:** A [calibrated] auxiliary reference standard as ***A*1** or ***A*2** (i.e. $i$ = **1** or **2**).

**β:** A constant (chosen number), but treated like a measured variable.

$C1_i$ Specified (***i**th***)** isotopic calibration constant for the $CO_2$ gas ***A*1** (i.e.: $C1_i = [{}^{A1}R_i/{}^{D}R_i] - 1$).

$C2_i$ ***i**th*** isotopic calibration constant for the $CO_2$ gas ***A*2** (i.e.: $C2_i = [{}^{A2}R_i/{}^{D}R_i] - 1$).

$\Delta_i$ ***i**th*** relative input/ experimental error ($\Delta_i = \frac{\Delta X_i}{X_i} = \frac{x_i - X_i}{X_i}$), e.g. $^{X}\Delta_J$, $^{Z1}\Delta_J$, $^{Z2}\Delta_J$ and $\Delta_\beta$ represent the errors in the estimates as**:** $x_{45/44}$, $z1_{45/44}$, $z2_{45/44}$ and **β**, respectively.

$Đ_d$ ***d**th*** relative COCP-output error ($Đ_d = \frac{dY_d}{Y_d} = \frac{y_d - Y_d}{Y_d}$). The 1st, 2nd and 3rd **stage**-output-errors are denoted as $^{I}Đ_d$ and $^{II}Đ_d$ (with**:** $d = i = J, K$) and $^{III}Đ_d$ ($d$ = **1**, **2**, **3**), respectively.

$E_d$**:** ***d**th*** constituent-elemental-isotopic abundance ratio in the sample $CO_2$ gas ***S***. $^{D}E_d$ refers to the "$E_d$ -value" in the [recommended/ desired] reference-*standard* $CO_2$ gas ***D***.

$\varepsilon_d$ ***d**th*** relative COCP-output uncertainty ($\varepsilon_d = |{}^{Max}Đ_d|$). $^{I}\varepsilon_d$, $^{II}\varepsilon_d$, and $^{III}\varepsilon_d$ refer to the 1st, 2nd and 3rd **stage** (i.e. 0th, 1st, and 2nd **cumulative**) output-uncertainties, respectively.

$F_i$ $F_i = u_i/{}^{G}u$ ($i$ = 1, 2 **…**). It enables the prediction of output-uncertainty ($\varepsilon_d$) even in a case where the uncertainty $u_i$ might vary with the variable to be measured.

$^{G}u$ Any **given** (i.e. preset *value* of) measurement-uncertainty $u_i$ (to be achieved).

$M_i^d$ *Relationship-sensitive-rate* of variation of ***d**th*-output with ***i**th*-input, cf. Eq. 9. Thus, $^{X}M_i^d$ and $^{Z1}M_i^d$ ($d = i$) represent the Eq. 6a specific rates of variation for $^{II}Y_d$ as a function of $X_i$ and $\boldsymbol{Z1_i}$, respectively. $^{III}\boldsymbol{M_i^d}$ ($\boldsymbol{d}$ = 1, 2, 3; and $\boldsymbol{i}$ = $\boldsymbol{J}$, $\boldsymbol{K}$, β) are the 3rd stage specific variation-rates. $\boldsymbol{M_i^d}$ ($\boldsymbol{i} = \boldsymbol{d}$) refers to a given Eq. 8; and so.

$\boldsymbol{R_i}$**:** $\boldsymbol{i^{th}}$ isotopic $CO_2$ abundance ratio in the *sample-gas* ($\boldsymbol{S}$); viz. $\boldsymbol{R_J} = \boldsymbol{R_{45/44}}$, and $\boldsymbol{R_K} = \boldsymbol{R_{46/44}}$. $\boldsymbol{r_i}$ is the estimate of $\boldsymbol{R_i}$. However the "$\boldsymbol{R_i}$-values", corresponding to the working reference $CO_2$ gas $\boldsymbol{W}$, the desired reference standard $CO_2$ gas $\boldsymbol{D}$, the auxiliary reference standards $CO_2$ gases $\boldsymbol{A1}$ and $\boldsymbol{A2}$, are referred to here as $^{W}\boldsymbol{R_i}$, $^{D}\boldsymbol{R_i}$, $^{A1}\boldsymbol{R_i}$ and $^{A2}\boldsymbol{R_i}$, respectively.

$\boldsymbol{u_i}$ $\boldsymbol{i^{th}}$ relative *measurement* [*input*] uncertainty ($\boldsymbol{u_i} = |^{\mathbf{Max}}\boldsymbol{\Delta_i}|$), e.g. $^{X}\boldsymbol{u_K}$ and $^{Z1}\boldsymbol{u_K}$ stand for uncertainties in the estimates of $\boldsymbol{X_{46/44}}$ and $\boldsymbol{Z1_{46/44}}$, respectively. Similarly, $\boldsymbol{u_\beta}$ refers the possible *uncertainty* in the *chosen value* of "β".

$\boldsymbol{[UF]_d}$ $\boldsymbol{d^{th}}$ COCP-uncertainty factor ($\boldsymbol{[UF]_d} = \boldsymbol{\varepsilon_d}/^{G}\boldsymbol{u}$). Actually, $\boldsymbol{[UF]_d}$ stands for "*collective COCP nature*" (cf. Eq. 13b/). $^{I}\boldsymbol{[UF]_d}$, $^{II}\boldsymbol{[UF]_d}$, and $^{III}\boldsymbol{[UF]_d}$ represent **1st**, **2nd** and **3rd** stage (i.e. 0th, 1st, and 2nd cumulative) uncertainty factors.

$\boldsymbol{X_i}$**:** Specified ($\boldsymbol{i^{th}}$) *measured/ input* variable (relating the sample $\boldsymbol{S}$)**:** $\boldsymbol{X_i} = ([\boldsymbol{R_i}/^{W}\boldsymbol{R_i}] - \mathbf{1})$. And $\boldsymbol{x_i}$ is the estimate of $\boldsymbol{X_i}$ ($\boldsymbol{i} = \boldsymbol{J}$ = 45/44 and $\boldsymbol{i} = \boldsymbol{K}$ = 46/44).

$\boldsymbol{Y_d}$**:** $\boldsymbol{d^{th}}$ COCP-*output* variable ($\boldsymbol{Y_d} = ([\boldsymbol{E_d}/^{D}\boldsymbol{E_d}] - \mathbf{1})$, with**:** $\boldsymbol{d}$ = 1, 2, 3). $\boldsymbol{y_d}$ is the estimate of $\boldsymbol{Y_d}$. $^{I}\boldsymbol{Y_d}$, $^{II}\boldsymbol{Y_d}$ and $^{III}\boldsymbol{Y_d}$ are the 1st, 2nd and 3rd stage outputs, respectively ($\{\boldsymbol{X_i}\} \rightarrow \{^{I}\boldsymbol{Y_d}\} \rightarrow \{^{II}\boldsymbol{Y_d}\} \rightarrow \{^{III}\boldsymbol{Y_d}\} \rightarrow \boldsymbol{Y_d(s)}$). Actually**:** $^{I}\boldsymbol{Y_d} = ([\boldsymbol{R_i}/^{D}\boldsymbol{R_i}] - \mathbf{1})$, with**:** $\boldsymbol{d} = \boldsymbol{i}$. However (by true values)**:** $^{II}\boldsymbol{Y_d} = \boldsymbol{R_i}$ (with**:** $\boldsymbol{d} = \boldsymbol{i}$); and**:** $^{III}\boldsymbol{Y_d} = \boldsymbol{E_d}$ (with**:** $\boldsymbol{d}$ = "13/12", "17/16" and "18/16").

$\boldsymbol{Z1_i}$ $\boldsymbol{i^{th}}$ *Input* [*measured*] variable in relation to the *auxiliary* standard $\boldsymbol{A1}$, i.e.**:** $\boldsymbol{Z1_i} = ([^{A1}\boldsymbol{R_i}/^{W}\boldsymbol{R_i}] - \mathbf{1})$; with**:** $\boldsymbol{i} = \boldsymbol{J}$ = 45/44 and**:** $\boldsymbol{i} = \boldsymbol{K}$ = 46/44.

$\boldsymbol{Z2_i}$ $\boldsymbol{i^{th}}$ *Measured* variable in relation to the *auxiliary* standard $\boldsymbol{A2}$, i.e.**:** $\boldsymbol{Z2_i} = ([^{A2}\boldsymbol{R_i}/^{W}\boldsymbol{R_i}] - \mathbf{1})$.

---

**Table 1.** Results by the standardization method/ COCP [as Eqs **6a**, **7** and **8**]: stage specific output estimates **[**and their errors (i) for *zero-error* in the input-data ($x_J$, $z1_J$, $x_K$, and $z1_K$, and 'β'), and (ii) for *varying* the inputs by exactly ±1%**]**

| → Estimate $y$ ↓ (Error Đ) Example No. | $^{II}y_J \times 10^3$ (% $^{II}$Đ$_J$) | $^{II}y_K \times 10^4$ (% $^{II}$Đ$_K$) | $^{III}y_1 \times 10^3$ (% $^{III}$Đ$_1$) | $^{III}y_2 \times 10^5$ (% $^{III}$Đ$_2$) | $^{III}y_3 \times 10^4$ (% $^{III}$Đ$_3$) | $y_1 \times 10^3$ (%Đ$_1$) | $y_2 \times 10^4$ (%Đ$_2$) | $y_3 \times 10^3$ (%Đ$_3$) |
|---|---|---|---|---|---|---|---|---|
| 0 *0 | | | | | | –11.484 **(0.0)** | | –21.310 **(0.0)** |
| 00 *00 | 11.857766 **(0.0)** | 40.546857 **(0.0)** | 11.108150 **(0.0)** | 37.480801 **(0.0)** | 20.231092 **(0.0)** | –11.484192 **(0.0)** | –107.124544 **(0.0)** | –21.310152 **(0.0)** |
| 1 *1 | 11.859413 (0.014) | 40.555382 (0.021) | 11.109798 (0.015) | 37.480750 (–0.00015) | 20.235348 (0.021) | –11.337488 (–1.3) | –107.139082 (0.014) | –21.104251 (–0.97) |
| 2 *2 | 11.856119 (–0.014) | 40.538333 (-0.021) | 11.106500 (–0.015) | 37.480935 (0.00036) | 20.226836 (-0.021) | –11.631027 (1.3) | –107.089245 (–0.033) | –21.516020 (0.97) |
| 3 *3 | 11.859413 (0.014) | 40.555382 (0.021) | 11.109636 (0.013) | 37.488734 (0.021) | 20.235339 (0.021) | –11.351705 (–1.15) | –105.030772 (–1.95) | –21.104666 (–0.96) |
| 4 *4 | 11.856119 (–0.014) | 40.538333 (–0.021) | 11.106663 (–0.013) | 37.472791 (–0.021) | 20.226845 (–0.021) | –11.616533 (1.15) | –109.238646 (1.97) | –21.515597 (0.96) |
| 5 *5 | 11.858647 (0.0074) | 40.548032 (0.0029) | 11.108940 (0.0071) | 37.485370 (0.012) | 20.231671 (0.0029) | –11.413901 (–0.61) | –105.918675 (–1.13) | –21.282110 (–0.13) |

***0** Results (as reported by Verkouteren and Lee,[9] i.e.) for**:** $X_J$ = –0.01055, $Z1_J$ = –0.00322, $X_K$ = **–**0.01182, $Z1_K$ = –0.00898, and β = 0.5.

***00** Results for **zero** input-errors (i.e. those obtained here by using the data-set referred to under example no. **0**).

***1** Results corresponding to the inputs as**:** $x_J$ = –0.0104445, $z1_J$ = –0.0032522, $x_K$ = –0.0117018, $z1_K$ = –0.0090698, and α = 0.5050 (i.e. for the input errors as**:** $^{X}\Delta_J = {}^{X}\Delta_K$ = –1%**,** and**:** $^{Z1}\Delta_J = {}^{Z1}\Delta_K = \Delta_\beta$ = 1%).

***2** Results for input data-errors as**:** $^{X}\Delta_J = {}^{X}\Delta_K$ = 1%, and**:** $^{Z1}\Delta_J = {}^{Z1}\Delta_K = \Delta_\beta$ = –1%.

***3** Results for input data-errors as**:** $^{X}\Delta_J = {}^{X}\Delta_K = \Delta_\beta$ = –1%, and**:** $^{Z1}\Delta_J = {}^{Z1}\Delta_K$ = 1%.

***4:** Results for input-errors as**:** $^{X}\Delta_J = {}^{X}\Delta_K = \Delta_\beta$ = 1%, and**:** $^{Z1}\Delta_J = {}^{Z1}\Delta_K$ = –1%.

***5:** Results for –1% errors in all input-data (i.e. for**:** $^{X}\Delta_J = {}^{X}\Delta_K = {}^{Z1}\Delta_J = {}^{Z1}\Delta_K = \Delta_\beta$ = –1%).

**Table 2.** Parameters ($\boldsymbol{M_i^d}$, $\boldsymbol{[UF]_d}$, and $\boldsymbol{\varepsilon_d}$) characterizing the different stages of the COCP represented by Eq. **6a**, Eq. **7** and Eq. **8**

| Cumulative Stage No. (Eq. No.) | **Out-put** ($\boldsymbol{Y_d}$) | **Rate of variation** (for $d^{th}$ output as a function of $i^{th}$ input: $\boldsymbol{M_i^d}$, cf. Eq. 9) | $\boldsymbol{d^{th}}$ **Uncertainty factor** ($\boldsymbol{[UF]_d}$, cf. Eq. 11, and/ or Eq. 13/ 13a/ 13b) | $\boldsymbol{d^{th}}$ **Uncertainty** ($\boldsymbol{\varepsilon_d}$, cf. Eq. 10, and/ or Eq. 12/ 12a/ 12b) |
|---|---|---|---|---|
| 1st (Eq. 6a) | $^{II}Y_J$ | $^{X}M_J^J = X_J / (X_J + 1) = -0.0107$<br>$^{Z1}M_J^J = -Z1_J / (Z1_J + 1) = 0.0032$ | $^{II}[UF]_J = \lvert {}^{X}M_J^J \rvert ({}^{X}u_J / {}^{G}u) + \lvert {}^{Z1}M_J^J \rvert ({}^{Z1}u_J / {}^{G}u)$<br>$= \lvert {}^{X}M_J^J \rvert {}^{X}F_J + \lvert {}^{Z1}M_J^J \rvert {}^{Z1}F_J$<br>$= 0.0139$ (with: $^{X}F_J = {}^{Z1}F_J = 1$) | $^{II}\varepsilon_J = (\lvert {}^{X}M_J^J \rvert \times {}^{X}u_J) + (\lvert {}^{Z1}M_J^J \rvert \times {}^{Z1}u_J) = {}^{II}[UF]_J\, {}^{G}u$<br>$= \mathbf{0.014}\,{}^{G}u$ |
| | $^{II}Y_K$ | $^{X}M_K^K = X_K / (X_K + 1) = -0.012$<br>$^{Z1}M_K^K = -Z1_K / (Z1_K + 1) = 0.0091$ | $^{II}[UF]_K = \lvert {}^{X}M_K^K \rvert ({}^{X}u_K / {}^{G}u) + \lvert {}^{Z1}M_K^K \rvert ({}^{Z1}u_K / {}^{G}u)$<br>$= \lvert {}^{X}M_K^K \rvert {}^{X}F_K + \lvert {}^{Z1}M_K^K \rvert {}^{Z1}F_K$<br>$= 0.0211$ (with: $^{X}F_K = {}^{Z1}F_K = 1$) | $^{II}\varepsilon_K = (\lvert {}^{X}M_K^K \rvert \times {}^{X}u_K) + (\lvert {}^{Z1}M_K^K \rvert \times {}^{Z1}u_K) = {}^{II}[UF]_K\, {}^{G}u$<br>$= \mathbf{0.021}\,{}^{G}u$ |
| 2nd (Eq. 7, i.e. here: Eqs. 2a, 3a, and 4a) | $^{III}Y_1$ | $^{III}M_J^1 = 1.07$, $^{III}M_K^1 = -0.0338$, and $^{III}M_L^1 = {}^{III}M_\beta^1 = 7.26\times10^{-4}$ | $^{III}[UF]_1 = (\lvert {}^{III}M_J^1 \rvert \times {}^{II}[UF]_J) + (\lvert {}^{III}M_K^1 \rvert \times {}^{II}[UF]_K) + (\lvert {}^{III}M_\beta^1 \rvert \times F_\beta) = 0.0163$ (with: $\{F_i = 1\}$) | $^{III}\varepsilon_1 = {}^{III}[UF]_1 \times {}^{G}u = \mathbf{0.016}\,{}^{G}u$ |
| | $^{III}Y_2$ | $^{III}M_J^2 = -0.0011$, $^{III}M_K^2 = 0.5005$, and $^{III}M_L^2 = {}^{III}M_\beta^2 = -0.0108$ | $^{III}[UF]_2 = (\lvert {}^{III}M_J^2 \rvert \times {}^{II}[UF]_J) + (\lvert {}^{III}M_K^2 \rvert \times {}^{II}[UF]_K) + (\lvert {}^{III}M_\beta^2 \rvert \times F_\beta) = 0.0213$ (with: $\{F_i = 1\}$) | $^{III}\varepsilon_2 = {}^{III}[UF]_2 \times {}^{G}u = \mathbf{0.021}\,{}^{G}u$ |
| | $^{III}Y_3$ | $^{III}M_J^3 = -0.0022$, $^{III}M_K^3 = 1.001$, and $^{III}M_L^3 = {}^{III}M_\beta^3 = 2.14\times10^{-5}$ | $^{III}[UF]_3 = (\lvert {}^{III}M_J^3 \rvert \times {}^{II}[UF]_J) + (\lvert {}^{III}M_K^3 \rvert \times {}^{II}[UF]_K) + (\lvert {}^{III}M_\beta^3 \rvert \times F_\beta) = 0.0211$ (with: $\{F_i = 1\}$) | $^{III}\varepsilon_3 = {}^{III}[UF]_3 \times {}^{G}u = \mathbf{0.021}\,{}^{G}u$ |
| 3rd (Eq. 8) | $Y_1$ | $M_1^1 = {}^{III}Y_1 / ({}^{III}Y_1 - {}^{D}E_1) = -86.1$ | $[UF]_1 = \lvert M_1^1 \rvert \times {}^{III}[UF]_1 = 1.40$ | $\varepsilon_1 = [UF]_1\, {}^{G}u = \mathbf{1.40}\,{}^{G}u$ |
| | $Y_2$ | $M_2^2 = {}^{III}Y_2 / ({}^{III}Y_2 - {}^{D}E_2) = -92.3$ | $[UF]_2 = \lvert M_2^2 \rvert \times {}^{III}[UF]_2 = 1.97$ | $\varepsilon_2 = [UF]_2\, {}^{G}u = \mathbf{1.97}\,{}^{G}u$ |
| | $Y_3$ | $M_3^3 = {}^{III}Y_3 / ({}^{III}Y_3 - {}^{D}E_3) = -45.9$ | $[UF]_3 = \lvert M_3^3 \rvert \times {}^{III}[UF]_3 = 0.97$ | $\varepsilon_3 = [UF]_3\, {}^{G}u = \mathbf{0.97}\,{}^{G}u$ |

**Table 3.** Results by the *conventional* normalization method or COCP (as Eqs. **14**, **15**, **16** and **17**): stage specific estimates as "*y*" and their errors as "**Đ**" **(**but corresponding to: (i) *zero* and (ii) exactly ±1.0% *errors* in the input-data [as: $x_J$, $z1_J$, $z2_J$, $x_K$, $z1_K$, $z2_K$ and, **β**]**)**

| **E**xample **No.** | ${}^{nI}y_J \times 10^3$ (% ${}^{nI}Đ_J$) | ${}^{nI}y_K \times 10^3$ (% ${}^{nI}Đ_K$) | ${}^{nII}y_J \times 10^3$ (% ${}^{nII}Đ_J$) | ${}^{nII}y_K \times 10^4$ (% ${}^{nII}Đ_K$) | ${}^{nIII}y_1 \times 10^3$ (% ${}^{nIII}Đ_1$) | ${}^{nIII}y_2 \times 10^5$ (% ${}^{nIII}Đ_2$) | ${}^{nIII}y_3 \times 10^4$ (% ${}^{nIII}Đ_3$) | ${}^{n}y_1 \times 10^3$ (% ${}^{n}Đ_1$) | ${}^{n}y_2 \times 10^4$ (% ${}^{n}Đ_2$) | ${}^{n}y_3 \times 10^3$ (% ${}^{n}Đ_3$) |
|---|---|---|---|---|---|---|---|---|---|---|
| 0 *0 | –14.391 **(0.0)** | –21.350 **(0.0)** | | | | | | –14.659 **(0.0)** | | –21.344 **(0.0)** |
| 00 *00 | –14.39119 **(0.0)** | –21.34952 **(0.0)** | 11.8223 **(0.0)** | 40.5453 **(0.0)** | 11.072708 **(0.0)** | 37.480199 **(0.0)** | 20.230442 **(0.0)** | –14.638176 **(0.0)** | –107.28333 **(0.0)** | –21.341569 **(0.0)** |
| 1 *1 | –14.69007 (2.077) | –21.58919 (1.123) | 11.8187 (–0.0303) | 40.5354 (–0.0245) | 11.069133 (–0.0322) | 37.479701 (–0.0013) | 20.225491 (–0.0245) | –14.956327 (2.17) | –107.41479 (0.123) | –21.581068 (1.122) |
| 2 *2 | –14.09822 (–2.036) | –21.11459 (–1.100) | 11.8258 (0.0297) | 40.5550 (0.0240) | 11.076212 (0.0316) | 37.480696 (0.0013) | 20.235295 (0.0240) | –14.326340 (–2.13) | –107.15216 (–0.122) | –21.106814 (–1.100) |
| 3 *3 | –14.69007 (2.077) | –21.58919 (1.123) | 11.8187 (–0.0303) | 40.5354 (–0.0245) | 11.069296 (–0.0308) | 37.471533 (–0.0231) | 20.225500 (–0.0244) | –14.941789 (2.07) | –109.57070 (2.13) | –21.580645 (1.120) |
| 4 *4 | –14.09822 (–2.036) | –21.11459 (–1.100) | 11.8258 (0.0297) | 40.5550 (0.0240) | 11.076052 (0.0302) | 37.488685 (0.0226) | 20.235286 (0.0239) | –14.340558 (–2.03) | –105.04359 (–2.09) | –21.107227 (–1.098) |
| 5 *5 | –14.39119 (0.0) | –21.34952 (0.0) | 11.8223 (0.0) | 40.5453 (0.0) | 11.072627 (–0.00073) | 37.484238 (0.0108) | 20.230438 ($-2.1\times10^{-5}$) | –14.645364 (0.049) | –106.21726 (–1.0) | –21.341778 (0.001) |

***0** Results as those reported by Verkouteren and Lee,[9] i.e. for**:** $X_J$ **= –**0.01055, $Z1_J$ = –0.00322, $Z2_J$ = –0.02881, $X_K$ **=** –0.01182, $Z1_K$ = –0.00898, $Z2_K$ = –0.0241, and **β** = 0.5.

***00** Results as obtained here (i.e. for**:** $x_i = [X_i + {}^{X}\Delta_i] = X_i$, $z1_i = [Z1_i + {}^{Z1}\Delta_i] = Z1_i$ **…** (***i*** **=** ***J*** and ***K***), and **β** = [**β** + $\Delta_\beta$] = **β**). The apparent variations of these results from those as the example no. "0" should be due to the difference in computational precision.)

***1** $x_J$ = –0.0106555, $z1_J$ = –0.0031878, $z2_J$ = –0.0285219, $x_K$ = –0.0119382, $z1_K$ = –0.0088902, $z2_K$ = –0.023859, and **β** = 0.4950 (i.e. results for data-errors as**:** ${}^{X}\Delta_J = {}^{X}\Delta_K$ = 1%, and ${}^{Z1}\Delta_J = {}^{Z1}\Delta_K = {}^{Z2}\Delta_J = {}^{Z2}\Delta_K = \Delta_\beta$ = –1%).

***2:** Results for data-errors as"${}^{X}\Delta_J = {}^{X}\Delta_K$ = –1%, and ${}^{Z1}\Delta_J = {}^{Z1}\Delta_K = {}^{Z2}\Delta_J = {}^{Z2}\Delta_K = \Delta_\beta$ = 1%".

***3:** ${}^{X}\Delta_J = {}^{X}\Delta_K = \Delta_\beta$ = 1%, and ${}^{Z1}\Delta_J = {}^{Z1}\Delta_K = {}^{Z2}\Delta_J = {}^{Z2}\Delta_K$ = –1%.

***4:** ${}^{X}\Delta_J = {}^{X}\Delta_K = \Delta_\beta$ = –1%, and ${}^{Z1}\Delta_J = {}^{Z1}\Delta_K = {}^{Z2}\Delta_J = {}^{Z2}\Delta_K$ = 1%.

***5:** ${}^{X}\Delta_J = {}^{X}\Delta_K = {}^{Z1}\Delta_J = {}^{Z1}\Delta_K = {}^{Z2}\Delta_J = {}^{Z2}\Delta_K = \Delta_\beta$ = –1%.

**Table 4.** Stage specific parameters ($M_i^d$, $[UF]_d$, and $\varepsilon_d$) for the COCP represented by **Eqs. 14**, **15**, **16** and **17**

| **C. Stage** (Eq. No.) | **Out-put** ($Y_d$) | **Rate of variation** (for $d^{th}$ output as a function of $i^{th}$ input**:** $M_i^d$, cf. Eq. 9) | $d^{th}$ **Uncertainty factor** ($[UF]_d$, cf. Eq. 11, and/ or Eq. 13/ 13a/ 13b) | $d^{th}$ **Uncertainty** ($\varepsilon_d$, cf. Eq. 10, and/or Eq. 12/ 12a/ 12b) |
|---|---|---|---|---|
| **0th** (Eq. 14) | ${}^{nI}Y_J$ | ${}^{nX}M_J^J = a_J X_J = 1.0280$ *1<br>${}^{nZ1}M_J^J = a_J Z1_J (X_J - Z2_J) / (Z2_J - Z1_J) = -0.2239$ *1<br>${}^{nZ2}M_J^J = a_J Z2_J (Z1_J - X_J) / (Z2_J - Z1_J) = -0.8041$ *1 | ${}^{nI}[UF]_J = (\lvert {}^{nX}M_J^J \rvert \times {}^{X}F_J) + (\lvert {}^{nZ1}M_J^J \rvert \times {}^{Z1}F_J) + (\lvert {}^{nZ2}M_J^J \rvert \times {}^{Z2}F_J) = (\lvert {}^{nX}M_J^J \rvert + \lvert {}^{nZ1}M_J^J \rvert + \lvert {}^{nZ2}M_J^J \rvert) = 2.056$ *2 | ${}^{nI}\varepsilon_J = {}^{nI}[UF]_J \times {}^{G}u$<br>$= \mathbf{2.056}\,{}^{G}u$ |
| | ${}^{nI}Y_K$ | ${}^{nX}M_K^K = a_K X_K = 0.5557$ *1<br>${}^{nZ1}M_K^K = a_K Z1_K (X_K - Z2_K) / (Z2_K - Z1_K) = -0.3429$ *1<br>${}^{nZ2}M_K^K = a_K Z2_K (Z1_K - X_K) / (Z2_K - Z1_K) = -0.2128$ *1 | ${}^{nI}[UF]_K = (\lvert {}^{nX}M_K^K \rvert \times {}^{X}F_K) + (\lvert {}^{nZ1}M_K^K \rvert \times {}^{Z1}F_K) + (\lvert {}^{nZ2}M_K^K \rvert \times {}^{Z2}F_K) = (\lvert {}^{nX}M_K^K \rvert + \lvert {}^{nZ1}M_K^K \rvert + \lvert {}^{nZ2}M_K^K \rvert) = 1.111$ *2 | ${}^{nI}\varepsilon_K = {}^{nI}[UF]_K \times {}^{G}u$<br>$= \mathbf{1.111}\,{}^{G}u$ |
| **1st** (Eq. 15) | ${}^{nII}Y_J$ | ${}^{nII}M_J^J = {}^{nI}Y_J / ({}^{nI}Y_J + 1) = -0.0146$ | ${}^{nII}[UF]_J = \lvert {}^{nII}M_J^J \rvert \times {}^{nI}[UF]_J = 0.030$ | ${}^{nII}\varepsilon_J = {}^{nII}[UF]_J \times {}^{G}u = \mathbf{0.030}\,{}^{G}u$ |
| | ${}^{nII}Y_K$ | ${}^{nII}M_K^K = {}^{nI}Y_K / ({}^{nI}Y_K + 1) = -0.0218$ | ${}^{nII}[UF]_K = \lvert {}^{nII}M_K^K \rvert \times {}^{nI}[UF]_K = 0.024$ | ${}^{nII}\varepsilon_K = {}^{nII}[UF]_K \times {}^{G}u = \mathbf{0.024}\,{}^{G}u$ |
| **2nd** (Eq. 16) | ${}^{nIII}Y_1$ | ${}^{nIII}M_J^1 = 1.07$, ${}^{nIII}M_K^1 = -0.034$, and ${}^{nIII}M_\beta^1 = 7.29\times10^{-4}$ | ${}^{nIII}[UF]_1 = 0.034$ | ${}^{nIII}\varepsilon_1 = \mathbf{0.034}\,{}^{G}u$ |
| | ${}^{nIII}Y_2$ | ${}^{nIII}M_J^2 = -0.0011$, ${}^{nIII}M_K^2 = 0.5005$, and ${}^{nIII}M_\beta^2 = -0.0108$ | ${}^{nIII}[UF]_2 = 0.023$ | ${}^{nIII}\varepsilon_2 = \mathbf{0.023}\,{}^{G}u$ |
| | ${}^{nIII}Y_3$ | ${}^{nIII}M_J^3 = -0.0022$, ${}^{nIII}M_K^3 = 1.001$, and ${}^{nIII}M_\beta^3 = 2.14\times10^{-5}$ | ${}^{nIII}[UF]_3 = 0.024$ | ${}^{nIII}\varepsilon_3 = \mathbf{0.024}\,{}^{G}u$ |
| **3rd** (Eq. 17) | ${}^{n}Y_1$ | ${}^{n}M_1^1 = {}^{nIII}Y_1 / ({}^{nIII}Y_1 - {}^{D}E_1) = -67.3$ | ${}^{n}[UF]_1 = \lvert {}^{n}M_1^1 \rvert \times {}^{nIII}[UF]_1 = 2.26$ | ${}^{n}\varepsilon_1 = {}^{n}[UF]_1 \times {}^{G}u = \mathbf{2.26}\,{}^{G}u$ |
| | ${}^{n}Y_2$ | ${}^{n}M_2^2 = {}^{nIII}Y_2 / ({}^{nIII}Y_2 - {}^{D}E_2) = -92.2$ | ${}^{n}[UF]_2 = \lvert {}^{n}M_2^2 \rvert \times {}^{nIII}[UF]_2 = 2.12$ | ${}^{n}\varepsilon_2 = {}^{n}[UF]_2 \times {}^{G}u = \mathbf{2.12}\,{}^{G}u$ |
| | ${}^{n}Y_3$ | ${}^{n}M_3^3 = {}^{nIII}Y_3 / ({}^{nIII}Y_3 - {}^{D}E_3) = -45.9$ | ${}^{n}[UF]_3 = \lvert {}^{n}M_3^3 \rvert \times {}^{nIII}[UF]_3 = 1.12$ | ${}^{n}\varepsilon_3 = {}^{n}[UF]_3 \times {}^{G}u = \mathbf{1.12}\,{}^{G}u$ |

*1 **Where:** $a_i = (C2_i - C1_i) / ([X_i - Z1_i] \times [C2_i - C1_i] + C1_i\,[Z2_i - Z1_i])$, $i = J$ and $K$.

*2 **For:** ${}^{X}F_i = {}^{Z1}F_i = {}^{Z2}F_i = 1$, i.e. **for:** ${}^{X}u_i = {}^{Z1}u_i = {}^{Z2}u_i = {}^{G}u$, $i = J$ and $K$.

**Table 5.** Results by the *expected method* of normalization (COCP as Eqs. **5b**, **6**, **7** and **8**)**:** variations in the stage specific outputs as "*y*" and/ or in their errors as "**Đ**" (for *zero-error*, and ±1.0% *errors*, in the input-data as**:** $\boldsymbol{x_J}$, $\boldsymbol{z1_J}$, $\boldsymbol{z2_J}$, $\boldsymbol{x_K}$, $\boldsymbol{z1_K}$, $\boldsymbol{z2_K}$ and, **β**)

| Example No. | $^{eI}y_J \times 10^3$ (% $^{eI}$Đ$_J$) | $^{eI}y_K \times 10^3$ (% $^{eI}$Đ$_K$) | $^{eII}y_J \times 10^3$ (% $^{eII}$Đ$_J$) | $^{eII}y_K \times 10^4$ (% $^{eII}$Đ$_K$) | $^{eIII}y_1 \times 10^3$ (% $^{eIII}$Đ$_1$) | $^{eIII}y_2 \times 10^5$ (% $^{eIII}$Đ$_2$) | $^{eIII}y_3 \times 10^4$ (% $^{eIII}$Đ$_3$) | $^{e}y_1 \times 10^3$ (% $^{e}$Đ$_1$) | $^{e}y_2 \times 10^4$ (% $^{e}$Đ$_2$) | $^{e}y_3 \times 10^3$ (% $^{e}$Đ$_3$) |
|---|---|---|---|---|---|---|---|---|---|---|
| 00 [*00] | 387.54993 (**0.0**) | –8.160074 (**0.0**) | 16.64357 (**0.0**) | 41.09173 (**0.0**) | 15.889259 (**0.0**) | 37.715474 (**0.0**) | 20.485225 (**0.0**) | 413.987402 (**0.0**) | –45.183657 (**0.0**) | –9.016316 (**0.0**) |
| 1 [*1] | 405.28731 (4.58) | –29.51136 (261.7) | 16.85633 (1.28) | 40.20715 (–2.15) | 16.110093 (1.39) | 37.311709 (–1.07) | 20.042768 (–2.16) | 433.639448 (4.747) | –151.75565 (235.9) | –30.420409 (237.4) |
| 2 [*2] | 370.25109 (–4.46) | 14.146369 (–273.4) | 16.43607 (–1.25) | 42.01588 (2.25) | 15.673248 (–1.36) | 38.141115 (1.13) | 20.947432 (2.26) | 394.764528 (–4.643) | 67.162105 (–248.6) | 13.343201 (–248.0) |
| 3 [*3] | 370.25109 (–4.46) | 14.146369 (–273.4) | 16.43607 (–1.25) | 42.01588 (2.25) | 15.673349 (–1.36) | 38.136067 (1.115) | 20.947440 (2.26) | 394.773513 (–4.641) | 65.829631 (–245.7) | 13.343575 (–248.0) |
| 4 [*4] | 405.28731 (4.58) | –29.51136 (261.7) | 16.85633 (1.28) | 40.20715 (–2.15) | 16.110363 (1.391) | 37.300201 (–1.10) | 20.042786 (–2.16) | 433.659930 (4.752) | –154.79308 (242.6) | –30.419533 (237.4) |
| 5 [*5] | 401.71503 (3.66) | 1.978347 (–124.2) | 16.81348 (1.021) | 41.51176 (1.022) | 16.055333 (1.05) | 37.907242 (0.51) | 20.694300 (1.02) | 428.766354 (3.57) | 5.432568 (–112.0) | 1.097793 (–112.2) |

[*00] Results for considering the true-values of input-**variables** (and the constants) to be as those reported in ref. [9], i.e. for**:** $X_J$ = –0.01055, $Z1_J$ = –0.00322, $Z2_J$ = –0.02881, $X_K$ = –0.01182, $Z1_K$ = –0.00898, $Z2_K$ = –0.0241**,** and **β** = 0.5 (see also Table 3 and/ or the text).

[*1] The input-*estimates* are as follows**:** $\boldsymbol{x_J = (X_J + {}^{X}\Delta_J) = (X_J - 1\%)}$ = –0.0104445, $\boldsymbol{z1_J = (Z1_J + {}^{Z1}\Delta_J) = (Z1_J + 1\%)}$ = –0.0032522, $\boldsymbol{z2_J = (Z2_J + {}^{Z2}\Delta_J) = (Z2_J - 1\%)}$ = –0.0285219, $\boldsymbol{x_K = (X_K + {}^{X}\Delta_K) = (X_K + 1\%)}$ = –0.0119382, $\boldsymbol{z1_K = (Z1_K + {}^{Z1}\Delta_K) = (Z1_K - 1\%)}$ = –0.0088902, $\boldsymbol{z2_K = (Z2_K + {}^{Z2}\Delta_K) = (Z2_K + 1\%)}$ = –0.024341, and $\boldsymbol{\beta = (\beta + \Delta_\beta) = (\beta - 1\%)}$ = 0.4950.

[*2] Input data-errors are varied as**:** $^{X}\Delta_J$ = 1%, $^{Z1}\Delta_J$ = –1%, $^{Z2}\Delta_J$ = 1%, $^{X}\Delta_K$ = –1%, $^{Z1}\Delta_K$ = 1%, $^{Z2}\Delta_K$ = –1%, and $\Delta_\beta$ = 1%.

[*3]**:** $^{X}\Delta_J = {}^{Z1}\Delta_K = {}^{Z2}\Delta_J$ = 1%, and $^{X}\Delta_K = {}^{Z1}\Delta_J = {}^{Z2}\Delta_K = \Delta_\beta$ = –1%.

[*4]**:** $^{X}\Delta_J = {}^{Z1}\Delta_K = {}^{Z2}\Delta_J$ = –1%, and $^{X}\Delta_K = {}^{Z1}\Delta_J = {}^{Z2}\Delta_K = \Delta_\beta$ = 1%.

[*5]**:** $^{X}\Delta_J = {}^{X}\Delta_K = {}^{Z1}\Delta_J = {}^{Z1}\Delta_K = {}^{Z2}\Delta_J = {}^{Z2}\Delta_K = \Delta_\beta$ = –1%.

**Table 6.** Parameters ($\boldsymbol{M_i^d}$, $\boldsymbol{[UF]_d}$, and $\boldsymbol{\varepsilon_d}$) of the COCP as Eq. **5b**, Eq. **6**, Eq. **7** and Eq. **8**

| **C. Stage** (Eq. No.) | **Out-put** ($\boldsymbol{Y_d}$) | **Rate of variation** (for $d^{th}$ output as a function of $i^{th}$ input: $\boldsymbol{M_i^d}$, cf. Eq. 9) | $\boldsymbol{d^{th}}$ **Uncertainty factor** ( $\boldsymbol{[UF]_d}$, cf. Eq. 11, and/ or Eq. 13/ 13a/ 13b) | $\boldsymbol{d^{th}}$ **Uncertainty** ($\boldsymbol{\varepsilon_d}$, cf. Eq. 10, and/or Eq. 12/ 12a/ 12b) |
|---|---|---|---|---|
| **0th** (Eq. 5b) | $^{I}Y_J$ | $^{eX}M_J^J = b_J \times X_J = -0.0382$ [*1]<br>$^{eZ1}M_J^J = b_J\, Z1_J\, (X_J + 1) / (Z2_J - Z1_J) = 0.4505$ [*1]<br>$^{eZ2}M_J^J = b_J\, Z2_J\, (X_J + 1) / (Z1_J - Z2_J) = -4.03$ [*1] | $^{eI}[UF]_J = (\lvert {}^{eX}M_J^J \rvert \times {}^{X}F_J) + (\lvert {}^{eZ1}M_J^J \rvert \times {}^{Z1}F_J) + (\lvert {}^{eZ2}M_J^J \rvert \times {}^{Z2}F_J) = (\lvert {}^{eX}M_J^J \rvert + \lvert {}^{eZ1}M_J^J \rvert + \lvert {}^{eZ2}M_J^J \rvert) = 4.5$ [*2] | $^{eI}\varepsilon_J = {}^{eI}[UF]_J \times {}^{G}u$<br>$= \mathbf{4.5}\,{}^{G}u$ |
| | $^{I}Y_K$ | $^{eX}M_K^K = b_K\, X_K = 1.45$ [*1]<br>$^{eZ1}M_K^K = b_K\, Z1_K\, (X_K + 1) / (Z2_K - Z1_K) = -72.2$ [*1]<br>$^{eZ2}M_K^K = b_K\, Z2_K\, (X_K + 1) / (Z1_K - Z2_K) = 193.7$ [*1] | $^{eI}[UF]_K = (\lvert {}^{eX}M_K^K \rvert \times {}^{X}F_K) + (\lvert {}^{eZ1}M_K^K \rvert \times {}^{Z1}F_K) + (\lvert {}^{eZ2}M_K^K \rvert \times {}^{Z2}F_K) = (\lvert {}^{eX}M_K^K \rvert + \lvert {}^{eZ1}M_K^K \rvert + \lvert {}^{eZ2}M_K^K \rvert) = 267.4$ [*2] | $^{eI}\varepsilon_K = {}^{eI}[UF]_K \times {}^{G}u$<br>$= \mathbf{267.4}\,{}^{G}u$ |
| **1st** (Eq. 6) | $^{II}Y_J$<br>$^{II}Y_K$ | $^{eII}M_J^J = {}^{I}Y_J / ({}^{I}Y_J + 1) = 0.2793$<br>$^{eII}M_K^K = {}^{I}Y_K / ({}^{I}Y_K + 1) = -0.008227$ | $^{eII}[UF]_J = \lvert {}^{eII}M_J^J \rvert \times {}^{eI}[UF]_J = 1.26$<br>$^{eII}[UF]_K = \lvert {}^{eII}M_K^K \rvert \times {}^{eI}[UF]_K = 2.20$ | $^{eII}\varepsilon_J = {}^{eII}[UF]_J \times {}^{G}u = \mathbf{1.260}\,{}^{G}u$<br>$^{eII}\varepsilon_K = {}^{eII}[UF]_K \times {}^{G}u = \mathbf{2.20}\,{}^{G}u$ |
| **2nd** (Eq. 7) | $^{III}Y_1$<br>$^{III}Y_2$<br>$^{III}Y_3$ | $^{eIII}M_J^1 = 1.05$, $^{eIII}M_K^1 = -0.024$, and $^{eIII}M_\beta^1 = 2.15\times10^{-4}$<br>$^{eIII}M_J^2 = -0.0015$, $^{eIII}M_K^2 = 0.5008$, and $^{eIII}M_\beta^2 = -0.0045$<br>$^{eIII}M_J^3 = -0.0031$, $^{eIII}M_K^3 = 1.002$, and $^{eIII}M_\beta^3 = 1.3\times10^{-5}$ | $^{eIII}[UF]_1 = 1.37$<br>$^{eIII}[UF]_2 = 1.11$<br>$^{eIII}[UF]_3 = 2.21$ | $^{eIII}\varepsilon_1 = \mathbf{1.37}\,{}^{G}u$<br>$^{eIII}\varepsilon_2 = \mathbf{1.11}\,{}^{G}u$<br>$^{eIII}\varepsilon_3 = \mathbf{2.21}\,{}^{G}u$ |
| **3rd** (Eq. 8) | $Y_1$<br>$Y_2$<br>$Y_3$ | $^{e}M_1^1 = {}^{III}Y_1 / ({}^{III}Y_1 - {}^{D}E_1) = 3.41$<br>$^{e}M_2^2 = {}^{III}Y_2 / ({}^{III}Y_2 - {}^{D}E_2) = -220.3$<br>$^{e}M_3^3 = {}^{III}Y_3 / ({}^{III}Y_3 - {}^{D}E_3) = -109.9$ | $^{e}[UF]_1 = \lvert {}^{e}M_1^1 \rvert \times {}^{eIII}[UF]_1 = 4.7$<br>$^{e}[UF]_2 = \lvert {}^{e}M_2^2 \rvert \times {}^{eIII}[UF]_2 = 244.1$<br>$^{e}[UF]_3 = \lvert {}^{e}M_3^3 \rvert \times {}^{eIII}[UF]_3 = 242.6$ | $^{e}\varepsilon_1 = {}^{e}[UF]_1 \times {}^{G}u = \mathbf{4.7}\,{}^{G}u$<br>$^{e}\varepsilon_2 = {}^{e}[UF]_2 \times {}^{G}u = \mathbf{244.1}\,{}^{G}u$<br>$^{e}\varepsilon_3 = {}^{e}[UF]_3 \times {}^{G}u = \mathbf{242.6}\,{}^{G}u$ |

[*1:] **Where:** $b_i = (C2_i - C1_i) / ([C2_i - C1_i] \times [X_i + 1] - [Z2_i - Z1_i])$, $i = J$ and $K$.
[*2:] **For:** $^{X}F_i = {}^{Z1}F_i = {}^{Z2}F_i = 1$, i.e. **for:** $^{X}u_i = {}^{Z1}u_i = {}^{Z2}u_i = {}^{G}u$, $i = J$ and $K$.

*a*) ***Standardization method*** **(use of single auxiliary standard: *A1*, see also Table 1)**

$$([x_J, z1_J]; [x_K, z1_K])^{Lab1} \rightarrow ([{}^{I}y_J]; [{}^{I}y_K]) \rightarrow ([{}^{II}y_J]; [{}^{II}y_K]) \xrightarrow{\beta + 0.5\%} ({}^{III}y_1,\ {}^{III}y_2,\ {}^{III}y_3) \rightarrow (y_1,\ y_2,\ y_3).$$

$\updownarrow$**1(% variation in labs-data)** $\updownarrow$1.2 $\updownarrow$≈1 $\updownarrow$.014 $\updownarrow$.021 $\updownarrow$(.015, .021, .021) $\updownarrow$(1.3, ≈2, 0.97)%

$$([x_J, z1_J]; [x_K, z1_K])^{Lab2} \rightarrow ([{}^{I}y_J]; [{}^{I}y_K]) \rightarrow ([{}^{II}y_J]; [{}^{II}y_K]) \xrightarrow{\beta - 0.5\%} ({}^{III}y_1,\ {}^{III}y_2,\ {}^{III}y_3) \rightarrow (y_1,\ y_2,\ y_3).$$

*b*) ***Conventional method*** **of using two auxiliary standards: *A1* and *A2*, cf. Table 3/ 4)**

$$([x_J, z1_J, z2_J];[x_K, z1_K, z2_K])^{Lab1} \rightarrow ([{}^{nI}y_J];[{}^{nI}y_K]) \rightarrow ([{}^{nII}y_J]; [{}^{nII}y_K]) \xrightarrow{\beta + 0.5\%} ({}^{nIII}y_1,\ {}^{nIII}y_2,\ {}^{nIII}y_3) \rightarrow ({}^{n}y_1,\ {}^{n}y_2,\ {}^{n}y_3).$$

$\updownarrow$**1(% variation in labs-data)** $\updownarrow$2.1 $\updownarrow$1.1 $\updownarrow$.03 $\updownarrow$.024 $\updownarrow$(.032, .023, .024) $\updownarrow$(2.2, 2.1, 1.1)%

$$([x_J, z1_J, z2_J];[x_K, z1_K, z2_K])^{Lab2} \rightarrow ([{}^{nI}y_J];[{}^{nI}y_K]) \rightarrow ([{}^{nII}y_J]; [{}^{nII}y_K]) \xrightarrow{\beta - 0.5\%} ({}^{nIII}y_1,\ {}^{nIII}y_2,\ {}^{nIII}y_3) \rightarrow ({}^{n}y_1,\ {}^{n}y_2,\ {}^{n}y_3).$$

*c*) ***Expected method*** **of employing two auxiliary standards: *A1* and *A2*, see alsoTable 5/ 6)**

$$([x_J, z1_J, z2_J];[x_K, z1_K, z2_K])^{Lab1} \rightarrow ([{}^{eI}y_J];[{}^{eI}y_K]) \rightarrow ([{}^{eII}y_J]; [{}^{eII}y_K]) \xrightarrow{\beta + 0.5\%} ({}^{eIII}y_1,\ {}^{eIII}y_2,\ {}^{eIII}y_3) \rightarrow ({}^{e}y_1,\ {}^{e}y_2,\ {}^{e}y_3).$$

$\updownarrow$**1(% variation in each data)** $\updownarrow$4.6 $\updownarrow$273 $\updownarrow$1.3 $\updownarrow$2.2 $\updownarrow$(1.4, 1.1, 2.3) $\updownarrow$(4.8, 249, 248)%

$$([x_J, z1_J, z2_J];[x_K, z1_K, z2_K])^{Lab2} \rightarrow ([{}^{eI}y_J];[{}^{eI}y_K]) \rightarrow ([{}^{eII}y_J]; [{}^{eII}y_K]) \xrightarrow{\beta - 0.5\%} ({}^{eIII}y_1,\ {}^{eIII}y_2,\ {}^{eIII}y_3) \rightarrow ({}^{e}y_1,\ {}^{e}y_2,\ {}^{e}y_3).$$

**Figure 1.** Output-comparability by different evaluation methods **[**in figure, $\boldsymbol{x_i}$, $\boldsymbol{z1_i}$ and $\boldsymbol{z2_i}$ stand for *measured* estimates of the *differential* isotopic-$CO_2$ abundance ratios $\boldsymbol{X_i}$, $\boldsymbol{Z1_i}$ and $\boldsymbol{Z2_i}$ (i.e. of ${}^{S/W}\delta_i^{CO_2}$, ${}^{A1/W}\delta_i^{CO_2}$ and ${}^{A2/W}\delta_i^{CO_2}$); ${}^{I}y_i$ and ${}^{II}y_i$ for estimates of the *scale-converted differential* and *absolute* isotopic $CO_2$ ratios ${}^{I}Y_i$ and ${}^{II}Y_i$ (i.e. of ${}^{S/D}\delta_i^{CO_2}$ and ${}^{S}R_i^{CO_2}$); respectively ($\boldsymbol{i}$ = $\boldsymbol{J}$ and $\boldsymbol{K}$). Similarly ${}^{III}y_1$, ${}^{III}y_2$ and ${}^{III}y_3$ refer to the estimates of ${}^{13/12}C$, ${}^{17/16}O$ and ${}^{18/16}O$ abundance ratios ${}^{III}Y_1$, ${}^{III}Y_2$ and ${}^{III}Y_3$ (i.e. of $E_1$, $E_2$ and $E_3$); and $y_1$, $y_2$ and $y_3$ are the estimates of desired *differential* ratios $Y_1$, $Y_2$ and $Y_3$ (i.e. of: ${}^{S/D}\delta_1$, ${}^{S/D}\delta_2$ and ${}^{S/D}\delta_3$); respectively. The pre-superscripts "***n***" and "***e***" are used to simply distinguish between *method-specific estimates* of any specified output-variable, cf. the text.**]**

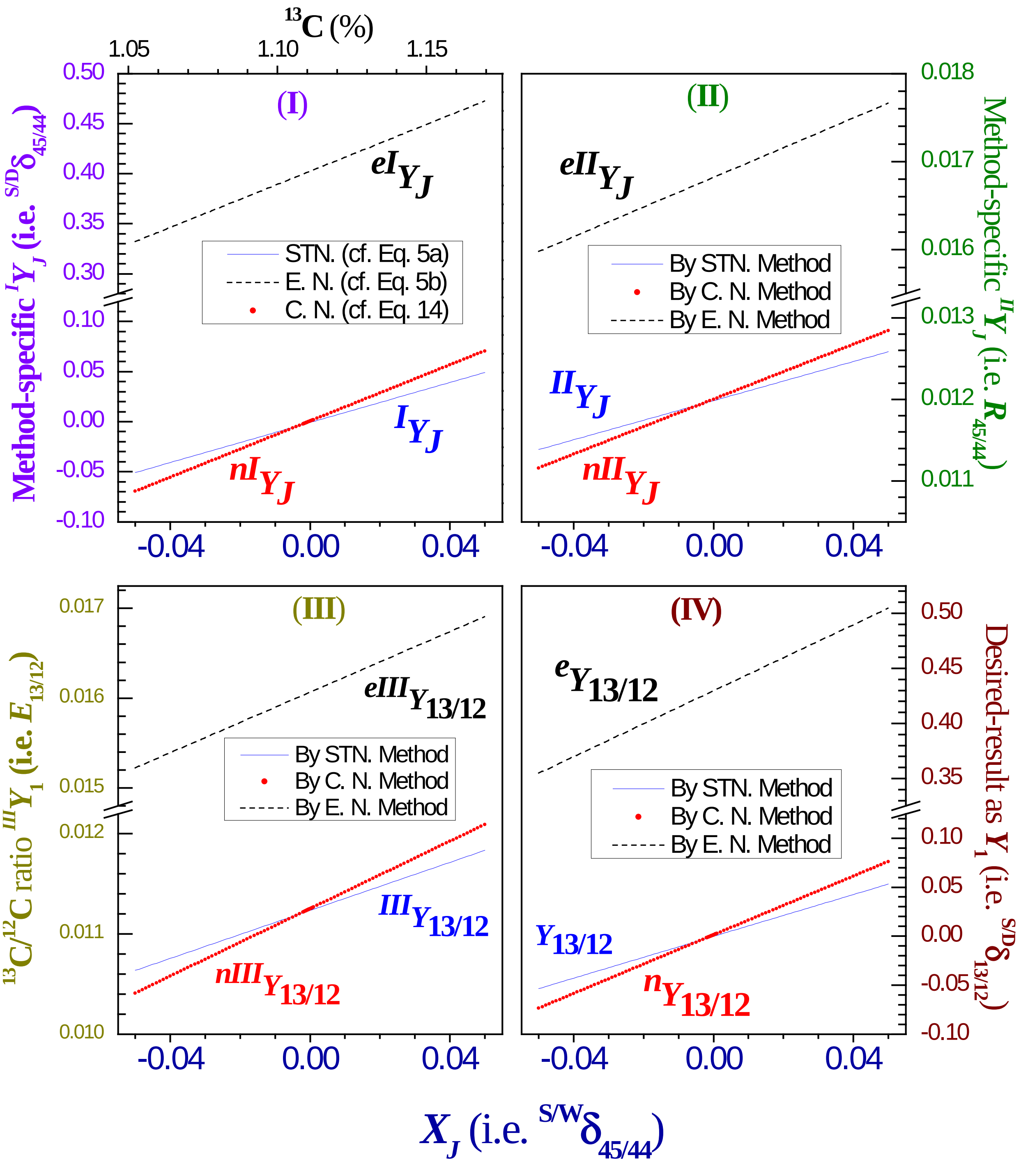

Figure 2. Variation of (I, II. III & IV-th) stage specific results (by the different methods**:** standardization "STN.", conventional normalization "C. N.", and expected normalization "E. N.") as a function of the measured variable $X_J$ (i.e. with $^{S/W}\delta_{45/44}$).

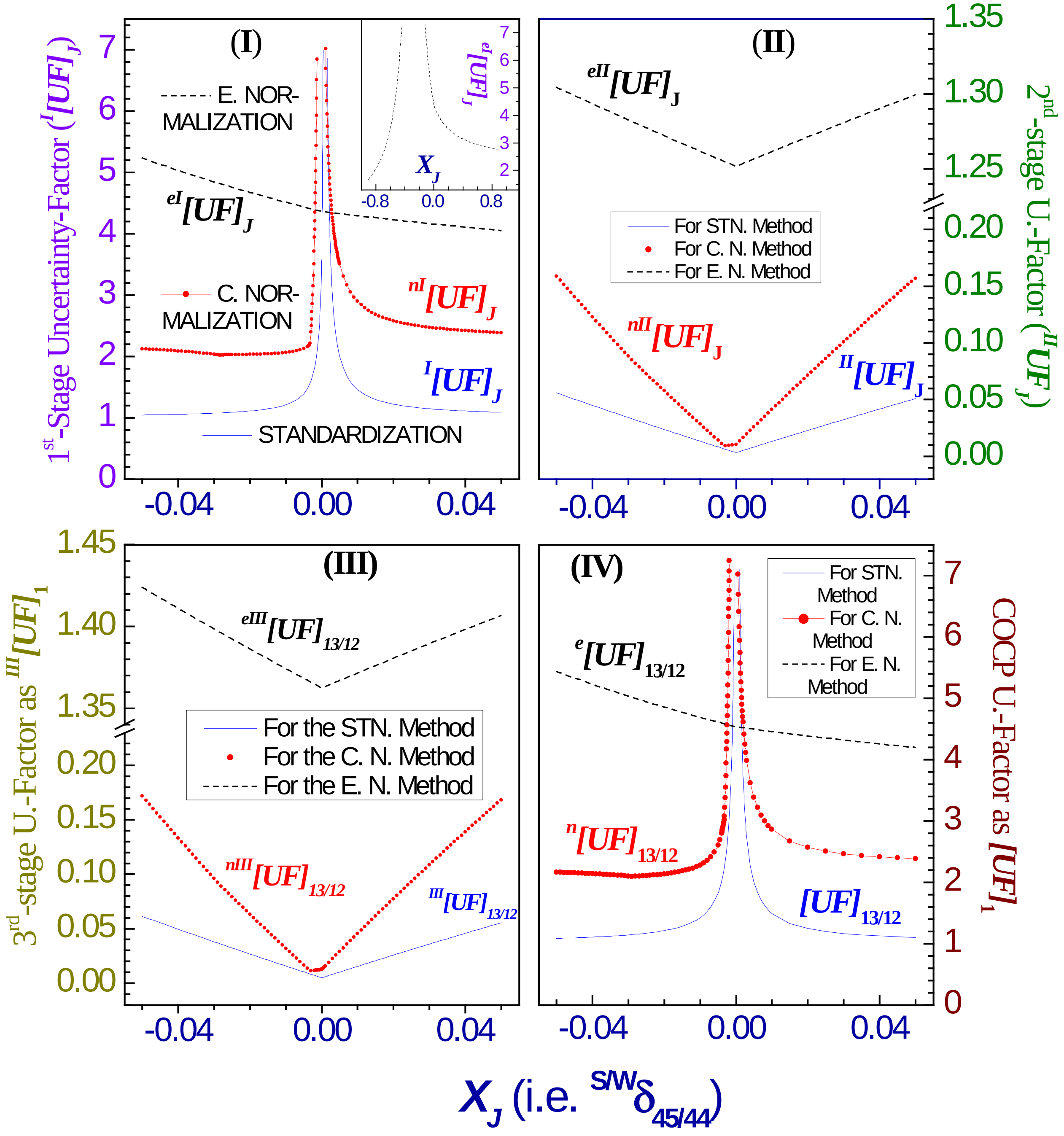

Figure 3. Variation of method- as well as stage-specific uncertainty-factors as a function of the measured variable $X_J$ (i.e. with $^{S/D}\delta_{45/44}$). (In figure, the *standardization*, *conventional* and *expected* normalization methods, are referred to as "STN", "C. N.", and "E. N.", respectively.)